\documentclass[11pt,letterpaper]{article}
\usepackage{amssymb}
\usepackage{mathtools}
\usepackage{axodraw2}
\usepackage{tikz}
\usepackage{tikz-feynman}
\usepackage{amscd}
\usepackage{amsmath}
\usepackage{slashed}
\usepackage{gensymb}
\usepackage{bm}
\usepackage{jheppub} 
\usepackage{booktabs}
\usepackage{array}  
\usepackage[capitalise]{cleveref}
\usepackage{comment}
\usetikzlibrary{shapes.misc}

\newcommand{\model}{Plan B Model}
\newcommand{\bsll}{$b \rightarrow s \ell^+ \ell^-$}

\title{Neutrino masses and mixing angles in the \model}

\author[a]{Ben Allanach}
\affiliation[a]{DAMTP, University of Cambridge, Wilberforce Road, Cambridge, 
  CB3 0WA, United Kingdom}
\emailAdd{ben.allanach.work@gmail.com}

\author{and Prabhoda Chandra Sarjapur\note{Corresponding author.}}
\emailAdd{prabhodacs@gmail.com}

\abstract{The Plan B Model was proposed to improve the global fit quality of the Standard Model to observables involving the $b\rightarrow s \ell^+ \ell^-$ transition whilst simultaneously providing a qualitative explanation for the small magnitudes of some off-diagonal quark mixing matrix entries. The model is based on applying an additional spontaneously broken $U(1)_{3B_3-L_e-2L_\mu}$ gauge symmetry to the Standard Model with family-dependent fermion charges. Right-handed neutrino fields are present and cancel gauge anomalies. We investigate the implications of the Plan B Model for neutrino masses and mixing angles. The model naturally suggests a Froggatt-Nielsen mechanism for neutrino masses and mixing hierarchies. We find that an implementation of the Froggatt-Nielsen mechanism works when additional suppression is present from a clockwork mechanism. We show how current neutrino oscillation data can be fit with order unity fundamental dimensionless couplings.}   


\begin{document} 
\maketitle
\flushbottom

\section{Introduction \label{sec:intro}}

Various measurements of $B$-meson decays at LHC experiments are in tension
with state-of-the-art Standard Model (SM) predictions, particularly when there is a bottom to
strange flavour transition accompanied by a di-lepton (i.e.\ di-electron or
di-muon) pair.
Measurements in several di-muon invariant mass-squared ($q^2$)
bins of the branching ratio of the $B_s$ meson decaying to a $\phi$ meson and
a di-muon pair
$BR(B_s \rightarrow \phi \mu^+ \mu^-)$ are $3.6\sigma$ smaller than      SM
predictions~\cite{LHCb:2021zwz,CDF:2012qwd}.
Some angular distributions in $B \rightarrow K^\ast \mu^+ \mu^-$
decays have been measured by LHC
experiments~\cite{LHCb:2013ghj,ATLAS:2018gqc,CMS:2017rzx,CMS:2015bcy,Bobeth:2017vxj,LHCb:2020gog}
to be several $\sigma$ short of state-of-the-art SM
predictions and
the same can be said of $BR(B \rightarrow K^\ast \mu^+
\mu^-)$~\cite{Parrott:2022zte}.
These tensions have not led to unambiguous claims of evidence of new physics
because of the difficulty in estimating the  SM
predictions and their theoretical uncertainties for these observables,
particularly as regards 
contributions coming from non-local 
corrections associated with a charm loop or
re-scattering through charmed mesons~\cite{Ciuchini:2021smi}. 
In any case, other measurements agree with the
predictions of the SM (to
within 1$\sigma$ or so), for example lepton flavour universality in $b \rightarrow s
\ell^+ \ell^-$ transitions\footnote{When we mention \bsll\ transitions, we
implicitly include the $CP$ conjugate mode $\bar b \rightarrow \bar s \ell^+
\ell^-$.}~\cite{LHCb:2022qnv} and SM predictions of $B_s-{\overline{B_s}}$ mixing~\cite{PDG}.
In Ref.~\cite{Allanach:2023uxz}, it was shown that the so-called \model\ can
significantly ameliorate a global fit of the SM to flavour observables via the predicted
family non-universal 
interactions of the model's TeV-scale $Z^\prime$ vector boson; the total $\chi^2$ (of
hundreds of observables) improved by 34 units in the model, which has only two
effective fit parameters. 
The best-fit point simultaneously
satisfies experimental constraints from di-lepton production at LEP2, which receives contributions
from an intermediate tree-level off-shell $Z^\prime$ boson.
Current LHC bounds upon
direct searches for the $Z^\prime$ are rather modest: $M_{Z^\prime} >
1.2$~TeV~\cite{Allanach:2024jls} for the central value of the global fit,
but there is essentially no meaningful LHC bound
in the parameter range that is compatible with both perturbativity and the
95$\%$ confidence level global fit~\cite{Allanach:2026yst} (one might require
$M_{Z^\prime}>0.25$ TeV so that the SMEFT approximation in the global fit
has a regime of validity). 
The \model\ augments the gauge group of the SM by a $U(1)_X$ factor. The $X$ charges of the SM Weyl fermionic quantum fields are assigned to be equal to thrice third family baryon number minus electron number minus twice muon number:
\begin{equation}
X = 3B_3-L_e-2L_\mu.    
\end{equation}
This is a vector-like charge assignment, i.e.\ a left-handed chiral fermionic
field has the same $X$ charge as its right-handed chiral counterpart. 
In order to allow a renormalisable top quark Yukawa coupling, the Higgs $X$
charge is fixed to vanish. This is necessary because the top quark Yukawa
coupling is the largest dimensionless parameter in the SM Lagrangian at the
electroweak scale (its value is around unity); as we shall see below, one can induce
smaller Yukawa couplings by banning them at the renormalisable level with a
symmetry but inducing them through spontaneous symmetry breaking. The
additional $U(1)_X$ gauge group factor is spontaneously broken by a SM-singlet
complex scalar 
field, the `flavon' $\theta$, which is assumed to have an $X$ charge of unity
and acquire a TeV-scale vacuum expectation value (VEV), leading to a TeV-scale
$Z^\prime$ vector boson with family-dependent gauge couplings. At the
renormalisable level, the  Yukawa couplings have the texture (where here and
later in the paper, $\times$ indicates a non-zero entry)
\begin{equation}
    Y_U , Y_D \sim \begin{pmatrix}
    \times & \times & 0 \\
        \times & \times & 0 \\      
        0 & 0 & \times
    \end{pmatrix}, \qquad
    Y_E \sim \begin{pmatrix}
    \times & 0 & 0 \\
        0& \times & 0 \\      
        0 & 0 & \times
    \end{pmatrix}. \label{chferms}
\end{equation}
At this renormalisable level, charged lepton flavour violation is then banned. 
Charged lepton flavour violation (CLFV) is heavily constrained by searches for it, especially in the $\mu \rightarrow e \gamma$ channel~\cite{MEGII:2023ltw}.
\cref{chferms} predicts zero values for the magnitudes of
Cabibbo-Kobayashi-Maskawa (CKM) matrix elements $|V_{ub}|,|V_{cb}|,|V_{td}|$
and $|V_{ts}|$, whereas $|V_{us}|$ and $|V_{cd}|$ are generically predicted to
be of order unity. The spontaneous breaking of $U(1)_X$  generically
predicts that
the CKM matrix elements are not strictly zero but merely much smaller than unity, a feature which agrees with measurements~\cite{PDG}. 

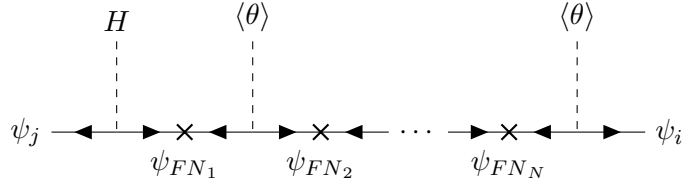
\begin{figure}[t] 
    \centering
    \begin{tikzpicture}
        \begin{feynman}
            \vertex (L) {$\psi_j$};
            \vertex [right=1.2cm of L] (F1);
            \vertex [right=0.9cm of F1] (G1);
            
            \vertex [right=0.9cm of G1] (F2);
            \vertex [right=0.9cm of F2] (G2);
            \vertex [right=0.9cm of G2] (dots) {$\cdots$};
            \vertex [below=0.1cm of G1] (FN1) {$\psi_{{FN}_1}$};
            \vertex [below=0.1cm of G2] (FN2) {$\psi_{{FN}_2}$};

            \vertex [right=1.2cm of dots] (GN);
            \vertex [right=2.1cm of dots] (FN);

            \vertex [below=0.1cm of GN] (FN3) {$\psi_{{FN}_N}$};       
            
            \vertex [right=0.9cm of FN] (R) {$\psi_i$};

            \vertex [above=1.2cm of F1] (phi1) {$H$};
            \vertex [above=1.2cm of F2] (phi2) {$\langle \theta \rangle$};
            \vertex [above=1.2cm of FN] (phiN) {$\langle \theta \rangle$};

            \diagram*{
            (F1) -- [fermion] (L),
                
            (F1)-- [fermion] (G1),
            (F2)-- [fermion] (G1),   
            
            (F2)-- [fermion] (G2),
            (dots)--[fermion] (G2),
            
            (dots)-- [fermion] (GN),
            (FN)-- [fermion] (GN),
                
            (FN)-- [fermion] (R),

            (phi1) -- [scalar] (F1),
            (phi2) -- [scalar] (F2),
            (phiN) -- [scalar] (FN),
            };
            \node [cross out, draw=black, thick, minimum size=5pt, inner sep=0pt] at (G1);
            \node [cross out, draw=black, thick, minimum size=5pt, inner sep=0pt] at (G2);
            \node [cross out, draw=black, thick, minimum size=5pt, inner sep=0pt] at (GN);
        \end{feynman}
    \end{tikzpicture}
    \caption{The ``spaghetti'' Feynman diagram leading to an approximate
      effective field theory FN operator $A\left(\frac{\langle\theta
        \rangle}{\Lambda_{\text{FN}}} \right)^{|X_i-X_j|} \overline{\psi_j} H
      \psi_i$. Each $\times$ represents a mass insertion of a different heavy
      beyond-the-SM fermionic field $\psi_{\text{FN}_i}$ whose charges are
      integers, differing by one unit in each case from left to
      right. $\psi_{\text{FN}_i}$ are in vector-like representations of the SM
      gauge group and have mass scale of order $\Lambda_{\text{FN}}$. $A$ is an
      effective coupling given by the product of all couplings in the diagram
      and another factor taking into account any mass factors between the $\psi_{\text{FN}_i}$ fields.}
    \label{fig:spag}
\end{figure}

The Froggatt-Nielsen (FN) mechanism~\cite{Froggatt:1978nt} constitutes a
suggestion as to where smaller Yukawa couplings could originate (following
$U(1)_X$ spontaneous symmetry breakdown). The idea is that there are some fermionic fields which are in vector-like representations of the SM gauge symmetry with masses above the scale $\langle \theta \rangle$ of $U(1)_X$ breaking.
Let us call this mass scale $\Lambda_{\text{FN}} \gg \langle \theta \rangle$. For the Yukawa couplings constrained to be zero at the renormalisable level, after $U(1)_X$ symmetry breaking, the dominant contribution to the effective Yukawa coupling term of the Lagrangian density between SM Weyl fermion $\psi_i$ and SM Weyl fermion $\psi_j$ and the Higgs doublet field $H$ is assumed to be 
\begin{equation}
    {\mathcal L}_{\overline{\psi_j} H \psi_i} = A \overline{\psi_j} H \psi_i
    \left| \frac{\langle \theta
      \rangle}{\Lambda_{\text{FN}}}\right|^{|X_j-X_i|} + H.c.,
    \label{effop}
\end{equation}
where $X_i$ is the $X$ charge of the Weyl fermionic field labelled by $i$ and $A$ is a product of (presumably) order unity dimensionless couplings of some underlying ultra-violet theory, of which we shall hear more below.  The number of powers of $\theta$ in the numerator fixes the Lagrangian density operator to be $U(1)_X$ invariant and, depending on $|X_i-X_j|$, provides varying levels of suppression of the effective Yukawa coupling. 
The idea is that the operator in (\ref{effop}) would be generated by underlying renormalisable physics in the ultra-violet, for example by the ``spaghetti diagram" in Fig.~\ref{fig:spag}.

The \model\  augments the SM Weyl fermionic field content by three
right-handed-neutrino-type fields, i.e.\ SM-gauge-singlet right-handed Weyl
fermionic fields~\cite{Allanach:2023uxz}. Such right-handed 
neutrinos have the potential to implement the seesaw
mechanism~\cite{Minkowski:1977sc,Yanagida:1980xy,Glashow:1979nm,Mohapatra:1979ia,Schechter:1979bn}, which suppresses Dirac neutrino masses by a
heavy neutrino mass scale in order to explain why neutrinos have such small
masses compared to the other observed SM fermionic fields via the
renormalisable Lagrangian density terms 
\begin{equation} \label{Neutrino Mass Lagrangian}
  {\mathcal L} = \ldots - {\overline L_i} (Y_\nu)_{ij} \tilde H \nu_{R_j} 
  - \frac{1}{2}
  {\overline{(\nu_{R_i})^c}} M_{ij} \nu_{R_j} + H.c., 
  \end{equation}
where\footnote{We denote the second Pauli matrix as $\sigma_2$.} $\tilde H=i \sigma_2 H^\ast$,
$i,j \in \{1,2,3\}$ are family indices (with an implied Einstein
summation convention on repeated indices), $SU(2)_L$ gauge indices have been
suppressed and ${}^c$ denotes charge conjugation. $(Y_\nu)_{ij}$ is a matrix
of dimensionless Yukawa couplings and 
$M_{ij}$ is a matrix of mass-dimension 1 Majorana masses. Our conventions for
the fields
can be found below in \cref{tab:lepton_charges}.

After spontaneous electroweak symmetry breaking, we can write the Higgs field
as $H = (0, (v+h)/\sqrt{2})^T$ (where $h$ is now a real scalar field and
$v\approx 246$~GeV
is the vacuum expectation value of the neutral Higgs field) and generate mass-like terms for the neutrinos through the Yukawa terms. Combining this with the Majorana terms and writing vectors in flavour space in bold font,
\begin{equation}
    \bm{\nu}_{L} =
    \begin{pmatrix}
        \nu_{eL} \\ \nu_{\mu L} \\ \nu_{\tau L}
    \end{pmatrix}, \qquad
    \bm{\nu}_{R} =
    \begin{pmatrix}
        \nu_{eR} \\ \nu_{\mu R} \\ \nu_{\tau R}
    \end{pmatrix},    
\end{equation}
we can rewrite the neutrino mass terms in \cref{Neutrino Mass Lagrangian} as 
\begin{equation}
    \mathcal{L}_\nu = - \frac{1}{2}\begin{pmatrix}
        \overline{\bm{\nu}_L} & \overline{(\bm{\nu}_R)^c}
    \end{pmatrix} \bm{M}_\nu \begin{pmatrix}
        {(\bm{\nu}_L)^c} \\ {\bm{\nu}_R}
    \end{pmatrix} + H.c.,
\end{equation}
where $\bm{M}_\nu$ is a $6 \times 6$ complex symmetric matrix
\begin{equation} \label{neutrino mass matrix}
    \bm{M}_\nu = \begin{pmatrix}
        0 & m_{\nu D} \\
        m_{\nu D}^T & M
    \end{pmatrix}
\end{equation}
and $m_{\nu D} = v\bm{Y}_\nu/\sqrt{2}$, where $\bm{Y}_\nu$ is the $3 \times 3$
matrix with entries $(Y_{\nu})_{ij}$. If the entries in the
right-handed Majorana mass matrix $M$ are much greater than
those in $m_{\nu D}$, which are set by 
the electroweak
scale $M_{ij} \gg v$, we can use block matrix perturbation theory to study the
system analytically. Then, the three
approximately right-handed neutrinos have a mass matrix $m_{\nu R} \approx M$
and the three  approximately left-handed neutrinos have a mass matrix $m_{\nu L}$~\cite{Gell-Mann:1979vob, Minkowski:1977sc,Glashow:1979nm}:
\begin{equation}
  {\mathcal L}_\nu \approx -\frac{1}{2} \left(\overline{{\bm \nu}_L} m_{\nu L} ({\bm
    \nu}_L)^c + \overline{({\bm \nu}_R)^c}m_{\nu R} {\bm \nu}_R \right)+ H.c.,
  \qquad m_{\nu L}\approx -m_{\nu D} M^{-1} m_{\nu D}^T. \label{mnu_eff}
\end{equation}
The suppression by $M^{-1}$ explains the lightness of the approximately
left-handed neutrino mass eigenstates, whereas the large mass scales present
in the approximately right-handed neutrino masses lead to heavy right-handed
neutrino mass eigenstates.  

In the \model, we already have right-handed neutrinos as well as a $U(1)_X$
symmetry that potentially implements the FN mechanism. It is natural then to
see if the FN mechanism can be made to work in a simple way that agrees with
measurements of neutrino oscillations and constraints upon neutrino mass in
detail utilising the Type I seesaw
mechanism~\cite{Minkowski:1977sc,Yanagida:1980xy,Glashow:1979nm,Mohapatra:1979ia,Schechter:1979bn}. \emph{This
is the central purpose of our paper}.
In order to present this analysis, we first
display the neutrino mass spectrum at the renormalisable level (in terms of
the SM plus right-handed neutrinos) in \cref{sec:origin}. Then we display
neutrino mass matrices in \cref{sec:offdiag}, analytically studying them
using perturbation theory in \cref{sec:anal}.
It is clear that the resulting neutrino masses are much too large when
compared to oscillation data. 
Thus, the model requires an additional mechanism to suppress the neutrino
masses. 
We achieve this with the
implementation of a clockwork mechanism
for the dimensionless neutrino Dirac Yukawa couplings, after which the  
mass spectrum can be obtained quite naturally. The IH case requires an additional 
suppression of two entries in the right-handed neutrino mass matrix. 
We summarise and conclude in \cref{sec:conc}.

\section{Origin of neutrino masses and lepton mixing} \label{sec:origin}

\begin{table}[ht]
\centering
\setlength{\tabcolsep}{10pt}     
\renewcommand{\arraystretch}{1.2}  
\begin{tabular}{|c|c|c|c|c|c|}
    \hline
     & $L_1, \,\, L_2, \,\, L_3$ & $e_1, \,\, e_2, \,\, e_3$ & $\nu_{R1}, \,\, \nu_{R2}, \,\, \nu_{R3}$& $H$ &$\theta$\\
    \hline
    $SU(3)$ & $\bm{1}$ & $\bm{1}$ & $\bm{1}$ & $\bm{1}$ & $\bm{1}$ \\
    $SU(2)$ & $\bm{2}$ & $\bm{1}$ & $\bm{1}$ & $\bm{2}$ & $\bm{1}$ \\
    $U(1)_Y$ & -1/2 & -1 & 0 & 1/2 & 0 \\
    $U(1)_X$ & $-1, \,\, -2, \,\, 0$
    & $-1, \,\, -2, \,\, 0$ & $-1, \,\, -2, \,\, 0$ & 0 & 1 \\
    \hline
\end{tabular}
\caption{Lepton and scalar field charge assignments under 
$SU(3)\!\times\!SU(2)\!\times\!U(1)_Y\!\times\!U(1)_X$. $L_i$ are left-handed
  Weyl fermion fields, whereas the rest of the fields are right-handed, and
  $H$ and $\theta$ are the complex scalar Higgs doublet and the complex scalar
  flavon, respectively.
}
\label{tab:lepton_charges}
\end{table}

While the primary motivation for the \model\  was to
change $B$-meson decay
observables via flavour changing interactions of a TeV-scale $Z^\prime$, the
model necessarily includes right-handed neutrino fields, making 
the neutrino sector of interest for study. 
In the model, the right-handed neutrinos are given
$X$ charges  $-1$, $-2$ and $0$, respectively, in order to cancel gauge
anomalies\footnote{Strictly speaking, the right-handed neutrino with $0$
$X$-charge is not required for gauge anomaly cancellation; it is added for
aesthetic reasons.}. As explained in \cref{sec:intro}, we wish to see if,  as
well as suppressing CLFV, the
$U(1)_X$ symmetry is compatible with neutrino masses and mixing parameters
which agree with those extracted from neutrino oscillation phenomena via the
FN mechanism. 

At the renormalisable level, ${Y}_\nu$ and ${M}$  are diagonal by virtue of them remaining invariant under the $U(1)_X$ charge assignment given in \cref{tab:lepton_charges}.
\begin{equation} \label{renormalizable Y and M}
    Y_{\nu}  \sim  
    \begin{pmatrix}
    \times & 0 & 0 \\
    0 & \times & 0 \\
    0 & 0 & \times 
    \end{pmatrix}, \,\,\,\,\,\,
    M  \sim 
    \begin{pmatrix}
    0 & 0 & 0 \\
    0 & 0 & 0 \\
    0 & 0 & \times 
    \end{pmatrix},
\end{equation}
which would imply no neutrino mixing and two massless right-handed neutrinos. These predictions are incompatible with experimental inference (given in \cref{tab:neutrino_angles_masses}) from neutrino oscillation phenomena~\cite{PDG}. In the basis where the charged lepton matrix is diagonal, which is natural for the \model, the Pontecorvo–Maki–Nakagawa–Sakata (PMNS) matrix arises solely due to the misalignment of the neutrino sector between the gauge and mass eigenbasis, i.e.
\begin{equation} \label{eqn:PMNSdefn1}
    \begin{pmatrix}
        \nu_{eL} \\ \nu_{\mu L} \\ \nu_{\tau L} 
    \end{pmatrix} = U_{PMNS} \begin{pmatrix}
        \nu_{1L} \\ \nu_{2 L} \\ \nu_{3 L} 
    \end{pmatrix},
\end{equation}
where 
\begin{equation} \label{eqn:PMNSdefn2}
    \text{diag}(m_1, m_2, m_3) = U_{PMNS}^\dagger m_{\nu L} U_{PMNS}^*.
\end{equation}
To explain the mass hierarchy and the PMNS matrix, one must go beyond the
renormalisable level of the Lagrangian and include
corrections from higher-dimension operators suppressed by inverse powers of
$\Lambda_\text{FN}$ in \cref{renormalizable Y and M}. The presence of such
higher-dimension operators is natural from the point of view of ultra-violet
(UV) completions of the model, and is the defining feature of FN models
\cite{Froggatt:1978nt}. 
\begin{table}[ht]
\centering
\setlength{\tabcolsep}{10pt}
\renewcommand{\arraystretch}{1.3}
\begin{tabular}{| l | c |c |} \hline
    \textbf{Parameter} & Normal Hierarchy (NH)& Inverted Hierarchy (IH) \\
    & (Best Fit $\pm 1\sigma$) & (Best Fit $\pm 1\sigma$) \\ \hline
    $\theta_{12}/^\circ$ & $33.68^{+0.73}_{-0.70}$ & $33.68^{+0.73}_{-0.70}$ \\
    $\theta_{23}/^\circ$ & $43.3^{+1.0}_{-0.8}$ & $47.9^{+0.7}_{-0.9}$ \\
    $\theta_{13}/^\circ$ & $8.56^{+0.11}_{-0.11}$ &  $8.59^{+0.11}_{-0.11}$  \\
    $\Delta m_{21}^2/10^{-5}\text{eV}^2$ & $7.49^{+0.19}_{-0.19}$ & $7.49^{+0.19}_{-0.19}$\\
    $\Delta m_{3l}^2/10^{-3}\text{eV}^2$ & $+2.513^{+0.021}_{-0.019}$ & $-2.484^{+0.020}_{-0.020}$\\
    \hline
\end{tabular}
\caption{Three-flavour neutrino oscillation {\tt NuFIT 6.0}~\cite{Esteban:2024eli} data as of September 2024 (for an alternative 
  global neutrino mass and mixing global fit, see Ref.~\cite{Capozzi:2025wyn}). 
  NH corresponds to $\Delta m_{3l}^2=\Delta m_{31}^2>0$ and  IH to $\Delta m_{3l}^2 = \Delta m_{32}^2<0$. 
The leptonic mixing angles (parameterising
the PMNS matrix) are defined in Eq.~(14.33) of Ref.~\cite{PDG}. The
uncertainties are all assumed to be Gaussian in the likelihood function.
\label{tab:neutrino_angles_masses}}
\end{table}

\subsection{Off-diagonal entries and the FN Mechanism} \label{sec:offdiag}

In the context of the Lagrangian terms leading to the neutrino mass matrix, the relevant FN terms are 
\begin{equation} \label{Yukawa Lagrangian FN}
    \small
    -\mathcal{L}_Y  \supset  \overline{\bm{L}_{L i}}  y_{ ij} \Big( \frac{\langle\theta\rangle}{\Lambda_{\text{FN}}}\Big)^{|X_i - X_j|} \tilde{H} \bm{\nu}_{R j}  +  \frac{\Lambda_{\text{FN}}}{2}  \overline{(\bm{\nu}_{Ri})^c} a_{ij}  \Big( \frac{\langle\theta\rangle}{\Lambda_{\text{FN}}}\Big)^{|X_i + X_j|} \bm{\nu}_{Rj} + H.c.,
\end{equation}
    where $ y_{ij}$ and $a_{ij}$ are assumed to be products of
$\mathcal{O}(1)$ coefficients and $\Lambda_{\text{FN}}$ sets the overall scale of the
right-handed Majorana masses.
Note that if
$X_i-X_j<0$, then we have $\theta^\dag$ insertions rather than insertions of
$\theta$ fields. 
Defining $\epsilon:=\langle\theta\rangle/\Lambda_{\text{FN}}$, the neutrino flavour matrices take the form 
\begin{equation} \label{FN Y and M}
     Y_{\nu} = 
    \begin{pmatrix}
         y_{11} &\epsilon  y_{12} & \epsilon  y_{13} \\
        \epsilon  y_{21}& y_{22}&\epsilon^{2} y_{23}\\
        \epsilon  y_{31}& \epsilon^{2} y_{32}& y_{33}
    \end{pmatrix}, \,\ 
    M =\Lambda_{\text{FN}}
    \begin{pmatrix}
    \epsilon^{2}a_{11}&\epsilon^{3}a_{12}&\epsilon a_{13}\\
    
    \epsilon^{3} a_{12} & \epsilon^{4} a_{22}& \epsilon^{2} a_{23}\\
    
    \epsilon a_{13} & \epsilon^{2}a_{23}&a_{33} 
    \end{pmatrix},
\end{equation}
where we are assuming $M_{33} =\Lambda_{\text{FN}} a_{33}$. The problem with
$M\sim \Lambda_{\text{FN}}$, however, is that since in our model
$\Lambda_{\text{FN}}\sim \mathcal{O}(1)\text{~TeV}$, we get the neutrino mass
to be on the order of  
\begin{equation}
    m_{\nu L} = -\frac{v^2}{2} Y_\nu M^{-1} Y_\nu^T \sim \frac{v^2}{\Lambda_{\text{FN}}} \sim 10 \,\text{GeV}.
\end{equation}
To match the current cosmological bounds $m_\nu \lesssim 0.1\text{~eV}$, we
require the Yukawa matrix in~\cref{FN Y and M} to be suppressed by
a universal factor of order $10^{-6}$. A natural mechanism to achieve this suppression,
without fine-tuning, is through the clockwork mechanism detailed
below~\cite{Kaplan:2015fuy,Giudice:2016yja,Choi:2015fiu,
  Park:2017yrn}. 

In this setup, we promote each fundamental right-handed (RH) neutrino $\nu_{Rj}$, where $j=1,2,3$ is the generation index, into a $K$-node chain of vector-like SM-singlet fermions. Let
$N_{Rj}^k \, (k=0,\dots,K)$ and $N_{Lj}^k \, (k=1,\dots,K)$ be the
singlet-Weyl fermions constituting this chain, all carrying the same $U(1)_X$ charge as the corresponding generation RH neutrino $\nu_{Rj}$ ($X_j = -1,-2,0$). The SM Higgs is then localised
at the $k=0$ site, such that the FN Dirac interaction with the
left-handed lepton doublets, $\overline{\bm{L}_{L i}}$, occurs exclusively
with $N_{R j}^0$. To ensure that the Majorana mass matrix remains at the energy scale $\Lambda_{\text{FN}}$, the bare Majorana insertions are tethered to the opposite boundary $k=K$ (the insertion at the other sites when expanded in terms of the physical field will be suppressed and hence can be ignored). The Lagrangian for the clockwork interaction sector is therefore,
\begin{equation} \label{eqn:CWInt}
    \small
    -\mathcal{L}_{\text{INT}}  \supset  \overline{\bm{L}_{L i}} y_{ ij} \Big( \frac{\langle\theta\rangle}{\Lambda_{\text{FN}}}\Big)^{|X_i - X_j|} \tilde{H} \bm{N}^0_{R j}  +  \frac{\Lambda_{\text{FN}}}{2}  \overline{(\bm{N}_{Ri}^K)^c} a_{ij}  \Big( \frac{\langle\theta\rangle}{\Lambda_{\text{FN}}}\Big)^{|X_i + X_j|} \bm{N}_{Rj}^K + H.c.,
\end{equation}
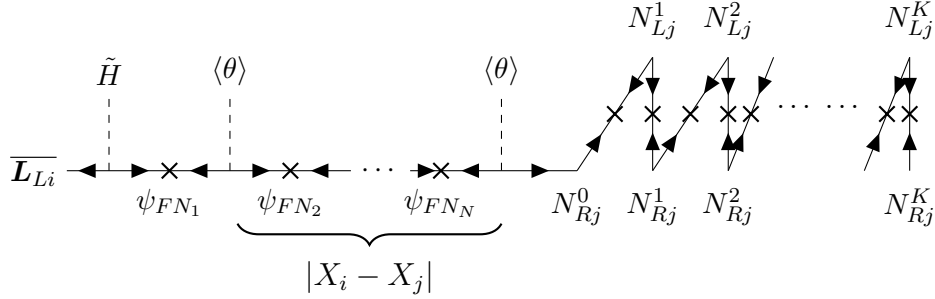
\begin{figure}[h!] 
    \centering
    \begin{tikzpicture}
        \begin{feynman}
            \vertex (L) {$\overline{\bm{L}_{L i}}$};
            
            \vertex[right=1.0cm of L] (V1);
            \vertex[right=0.8cm of V1] (I1);

            \vertex[right=0.8cm of I1] (V2);
            \vertex[right=0.8cm of V2] (I2);

            \vertex[right=0.8cm of I2] (dots1) {$\cdots$};

            \vertex[right=0.8cm of dots1] (IN);
            \vertex[right=0.8cm of IN] (VN);

            \vertex[right=1.0cm of VN] (N0);
            \vertex[below=0.1cm of N0] (N0n){$N_{Rj}^0$};
            
            \vertex[right=1.0cm of N0] (NR1);
            \vertex[below=0.1cm of NR1] (NR1n){$N_{Rj}^1$};
            
            \vertex[above=1.5cm of NR1] (NL1);
            \vertex[above=0.1cm of NL1] (NL1n){$N_{Lj}^1$};
            
            \vertex[right=1.0cm of NR1] (NR2);
            \vertex[below=0.1cm of NR2] (NR2n){$N_{Rj}^2$};
            
            \vertex[right=1.0cm of NL1] (NL2);
            \vertex[above=0.1cm of NL2] (NL2n){$N_{Lj}^2$};
            
            \vertex[right=0.6cm of NR2] (NRm);
            \vertex[above=0.7cm of NR2] (M2){$\quad  \quad \quad \quad \,\, \quad \quad \dots \,\, \dots$};
            \vertex[right=0.6cm of NL2] (NLm);
            
            \vertex[right=1.8cm of NLm] (NLK);
            \vertex[above=0.1cm of NLK] (NLKn){$N_{Lj}^K$};
            
            \vertex[below=1.5cm of NLK] (NRK);
            \vertex[below=0.1cm of NRK] (NRKn){$N_{Rj}^K$};
            
            \vertex[above=0.3cm of NRK] (MK);

            \vertex[left=0.6cm of NRK] (NRKm);
            \vertex[right=1.8cm of M2] (MKm);
            \vertex[left=1.0cm of NLK] (NLKm);

            \vertex[below=0.1cm of I1] (psi1) {$\psi_{FN_1}$};
            \vertex[below=0.1cm of I2] (psi2) {$\psi_{FN_2}$};
            \vertex[below=0.1cm of IN] (psiN) {$\psi_{FN_N}$};
            \vertex[above=1.0cm of V1] (phi1) {$\tilde{H}$};
            \vertex[above=1.0cm of V2] (phi2) {$\langle \theta \rangle$};
            \vertex[above=1.0cm of VN] (phiN) {$\langle \theta \rangle$};

            \path (N0) -- (NL1) coordinate[midway] (C1);
            \path (NL1) -- (NR1) coordinate[midway] (C2);
            \path (NR1) -- (NL2) coordinate[midway] (C3);
            \path (NL2) -- (NR2) coordinate[midway] (C4);
            \path (NR2) -- (NLm) coordinate[midway] (C5);
            \path (NRKm) -- (NLK) coordinate[midway] (CK1);
            \path (NLK) -- (NRK) coordinate[midway] (CK2);
            
            \diagram*{
            (V1)--[fermion] (L);
            (V1)--[fermion] (I1);
            (V2)-- [fermion] (I1);
            (V2) --[fermion] (I2);
            (dots1)--[fermion] (I2);
            (dots1) --[fermion] (IN);
            (VN) -- [fermion] (IN);
            (VN) -- [fermion] (N0);

            (N0) --[fermion] (C1);
            (NL1) --[fermion] (C1);
            
            (NL1) --[fermion] (C2);
            (NR1) --[fermion] (C2);
            
            (NR1) --[fermion] (C3);
            (NL2) --[fermion] (C3);
            
            (NL2) --[fermion] (C4);
            (NR2) --[fermion] (C4);
            
            (NR2) --[fermion] (C5);
            (NLm) --[fermion] (C5);
            
            (NRKm) --[fermion] (CK1);
            (NLK) --[fermion] (CK1);
            
            (NLK) --[fermion] (CK2);
            (NRK) --[fermion] (CK2);

            (V1) -- [scalar] (phi1);
            (V2) -- [scalar] (phi2);
            (VN) -- [scalar] (phiN);
            };

            \node [cross out, draw=black, thick, minimum size=5pt, inner sep=0pt] at (I1) {};
            \node [cross out, draw=black, thick, minimum size=5pt, inner sep=0pt] at (I2) {};
            \node [cross out, draw=black, thick, minimum size=5pt, inner sep=0pt] at (IN) {};
            
            \node [cross out, draw=black, thick, minimum size=5pt, inner sep=0pt] at (C1) {};
            \node [cross out, draw=black, thick, minimum size=5pt, inner sep=0pt] at (C2) {};
            \node [cross out, draw=black, thick, minimum size=5pt, inner sep=0pt] at (C3) {};
            \node [cross out, draw=black, thick, minimum size=5pt, inner sep=0pt] at (C4) {};
            \node [cross out, draw=black, thick, minimum size=5pt, inner sep=0pt] at (C5) {};
            \node [cross out, draw=black, thick, minimum size=5pt, inner sep=0pt] at (CK1) {};
            \node [cross out, draw=black, thick, minimum size=5pt, inner sep=0pt] at (CK2) {};

            \draw [decorate, decoration={brace, amplitude=8pt, mirror}, thick] 
            ([xshift=0.1cm, yshift=-0.7cm]V2.center) -- ([yshift=-0.7cm]VN.center) 
            node [midway, below=10pt] {\large $|X_i-X_j|$};
        \end{feynman}
    \end{tikzpicture}
    \caption{A clockworked FN mechanism. The FN part of the mechanism is on
      the left-hand side of the diagram and the clockwork part on the
      right-hand side.}
    \label{fig:CW with FN}
\end{figure}
By introducing $K+1$ RH fermions and $K$ LH fermions per generation in the chain, the chiral symmetry group $U(1)_R^{K+1}\times U(1)_L^K$ is preserved. To break all but one chiral symmetry group, we add $K$ mass terms, $\mu \gtrsim \Lambda_{\text{FN}}$, on each of the nodes, and a series of nearest neighbour mass terms, $q_Y\mu$ (where $q_Y>1$), between the nodes,
\begin{equation}
    \mathcal{L}_{\text{CW}} = \mathcal{L}_{\text{KIN}} - \mu \sum_{k=1}^{K} (\overline N_{Lj}^{k}  N_{Rj}^{k} - q_Y \overline N_{L j}^{k}  N_{Rj}^{k-1}) + H.c.
\end{equation}
The chain only couples to $\overline{\bm{L}_{L_i}} \tilde H N_{R j }^0$,
because the Higgs doublet and FN mediators are restricted to site 0 --- like
the lepton-number violating spurion to site $K$ --- by the assignment of
fields to sites with site-local interactions. The unbroken symmetry group now
guarantees the existence of a massless zero mode. The $K\times(K+1)$ mass
matrix is, 
\begin{equation}
    \mathcal{M}_N = \mu 
    \begin{pmatrix}
        -q_Y & 1 & 0 & \dots & 0 \\
        0 & -q_Y & 1 & \dots & 0 \\
        \vdots & \ddots & \ddots & \ddots & 0 \\
        0 & \dots & 0 & -q_Y & 1
    \end{pmatrix},
\end{equation}
and is diagonalised by unitary rotations, $\text{diag}(0,M_1,\dots,M_K) = (V^L)^T\mathcal{M}_N V^R$. This gives one zero mode, to be identified with the RH neutrino $\nu_{R j}$. Because this zero mode is exponentially localised at the $k=K$ boundary, its overlap with the bare field at the SM boundary ($k=0$) is highly suppressed. Expanding the bare field $N_{Rj}^0$ in terms of the physical basis yields:
\begin{equation}
    N_{R j}^0 = V_{00}^R \nu_{R j} + \text{heavy modes} = \sqrt{\frac{q_Y^2-1}{q_Y^{2} - q_Y^{-2K}}}\frac{1}{q_Y^K} \nu_{R j} + \dots \approx q_Y^{-K} \nu_{R j},
\end{equation}
where the last approximation is in the regime $q_Y\gg 1,\ K \geq 1$. This gives us the exponential suppression as promised. Therefore~\cref{eqn:CWInt} becomes,
\begin{equation} \label{eqn:eft Yukawa and Majorana}
    -\mathcal{L}_{\text{EFT}} = q_Y^{-K}  y_{ij} \overline{\bm{L}_{L i}} \bigg(\frac{\langle \theta\rangle}{\Lambda_{\text{FN}}} \bigg)^{|X_i-X_j|} \tilde{H} \nu_{R j} +\frac{\Lambda_{\text{FN}}}{2}  \overline{(\bm{\nu}_{Ri})^c} a_{ij}  \Big( \frac{\langle\theta\rangle}{\Lambda_{\text{FN}}}\Big)^{|X_i + X_j|} \bm{\nu}_{Rj} +H.c.
\end{equation}
We require the same values of $q_Y$ and $K$ for all three chains. 
For $q_Y = 10, K=6$, we get the suppression we need (or, for example, one
could use $q_Y=3, K=13$ to achieve the same suppression factor if one worries
that $q_Y=10$ is itself a fundamental dimensionless coupling of order 10). The physical
active neutrino mass matrix is given by the Type I seesaw formula: 
\begin{equation} \label{eqn:mnu_effective}
    m_{\nu L} = -q_Y^{-2K}\frac{v^2}{2} Y_\nu M^{-1} Y_\nu^T.
\end{equation} 
\cref{FN Y and M} along with \cref{eqn:mnu_effective} delivers the texture of the neutrino Yukawa and Majorana
mass matrices in the \model.

Experimental CLFV constraints imply that the diagonal texture in
\cref{chferms} must be valid to a much better approximation than in the
neutrino Yukawa matrix. This could easily be enforced in the field theory
which is effective just above $\Lambda_\text{FN}$ by omitting vector-like (with
respect to the SM gauge group) representations of charged leptons, thus
possessing no tree-level spaghetti diagrams which contribute to $Y_E$. With
this in mind, we shall henceforth neglect contributions from charged-lepton
mixing in the gauge eigenbasis, therefore determining the PMNS matrix by use
of  \cref{eqn:PMNSdefn2}.
Because of the current lack of sizeable experimental constraints on CP violation in the leptonic sector, we neglect phases in $y_{ij}$ and $a_{ij}$. We are thus fitting to a scheme where all complex phases in the PMNS matrix in Eq.~(14.33) of Ref.~\cite{PDG} are zero.
Complex phases could be analysed at a later date once experimental
constraints on leptonic $CP$ violation have sharpened.
We explore whether \cref{FN Y and M} suffices to explain the values of the neutrino mass hierarchies and the values of the PMNS mixing angles in the following section.

\subsection{Analytic expansion of masses and mixing angles for NH} \label{sec:anal}
\cref{neutrino mass matrix} makes it clear that we shall need to calculate
$M^{-1}$ in order to calculate $m_{\nu_L}$, which we need to work out the
neutrino phenomenology. 
For this, we find it useful to further factorise $M$ from \cref{FN Y and M}:
\begin{equation} \label{eqn:MajoranaMatrixFact}
M = \Lambda_{\text{FN}} \Sigma A \Sigma, \qquad 
    \Sigma = \begin{pmatrix}
    \epsilon^{1} & 0 & 0 \\
     0 & \epsilon^2 & 0 \\
     0 & 0 & 1 \\
    \end{pmatrix}, \qquad
A =     \begin{pmatrix}
        a_{11} & a_{12}  & a_{13} \\
        a_{12} & a_{22}  & a_{23} \\
        a_{13} & a_{23} & a_{33} 
    \end{pmatrix}.
\end{equation}
This implies that the inverse of the Majorana mass matrix can also be factorised:
\begin{equation}
     M^{-1} =\frac{\Sigma^{-1} A^{-1} \Sigma^{-1}}{\Lambda_{FN}} = \frac{1}{\Lambda_{\text{FN}} }\begin{pmatrix}
     \epsilon ^{-1} & 0 & 0 \\ 
     0 & \epsilon ^{-2} & 0 \\ 
     0 & 0 & 1 
     \end{pmatrix}
     \begin{pmatrix}
        a_{11} & a_{12}  & a_{13} \\
        a_{12} & a_{22}  & a_{23} \\
        a_{13} & a_{23} & a_{33} 
    \end{pmatrix}^{-1}
    \begin{pmatrix}
     \epsilon ^{-1} & 0 & 0 \\ 
     0 & \epsilon ^{-2} & 0 \\ 
     0 & 0 & 1 
     \end{pmatrix}.
\end{equation}
We denote the entries of $A^{-1}$ as $\alpha_{ij}$. 
Each $\alpha_{ij}$ is the ratio of a real homogeneous quadratic of the
$a_{ij}$ and a real homogeneous cubic of the $a_{ij}$. If the $a_{ij}$ are of
${\mathcal O}(1)$, one might naively expect the $\alpha_{ij}$ to be of ${\mathcal O}(1)$ too.
However, Ref.~\cite{Allanach:2026rkb} points out that 
${\mathcal O}(1)$ requires a broader
definition than one might naively think. 
It models the ${\mathcal O}(1)$ dimensionless coupling expectation with 
normal probability distributions of width unity and
centred on zero. One can then measure the adherence (or otherwise) of the ${\mathcal O}(1)$ expectation\footnote{Here, we refer to fundamental Yukawa couplings as
those that are supposed to be truly of order unity.} by using the spread $r$:
the ratio of the maximum magnitude of dimensionless coupling divided 
by the minimum magnitude of dimensionless coupling, where $r \geq 1$. The central result of
Ref.~\cite{Allanach:2026rkb} is that, even though extreme values of the
couplings are exponentially suppressed, the tail of the spread is not:
it is typically only suppressed $\propto 1/r^2$ in its probability
density functions (PDFs) and the chance of 
the spread being larger than 
a fixed value only becomes larger with more dimensionless order one couplings.  
For a theory with 9 independent identically-distributed unit normal couplings, the probability that the spread is greater than 30 is 0.35, for example.
In
what follows, we shall be dealing with some tens of fundamental dimensionless
couplings that we may assume are ${\mathcal O}(1)$. We may then expect
that, starting from ${\mathcal O}(1)$ expectations for dimensionless couplings, one or two of the ratios of
independently-distributed ${\mathcal O}(1)$ parameters may differ by factors of a
few tens.
Here, we shall follow 
Ref.~\cite{Allanach:2026rkb} and quantify naturalness according to the spread
\begin{equation}
  r:=|Y_{\text{max}}|/|Y_{\text{min}}|, \label{eq:spread}
\end{equation}
  i.e.\ the
ratio of the maximum fundamental coupling to the
minimum fundamental coupling in absolute value.
We shall then view $r$ less than a few tens as natural; to be definite,
we take $r \leq 30$ to be the default.

The light approximately-left-handed neutrino mass matrix is given at tree level by the seesaw formula \cref{mnu_eff}.
Factoring some overall powers of $\epsilon$ in the inverse Majorana matrix
(see discussion below), we expand $m_\nu:=-m_{\nu_L}$ (the minus sign can be
absorbed by field re-definitions)
as a series in $\epsilon$:
\begin{equation}  \label{eqn:mnu_expansion}
    m_\nu = q_Y^{-2K}\frac{v^2}{2\Lambda_{\text{FN}}} \epsilon^{-4} \left[X^{(0)} + \epsilon X^{(1)} + \epsilon^2 X^{(2)} + \mathcal{O}(\epsilon^3)\right], 
\end{equation}
where:
\begin{equation} \label{Lowest order - Unchanged}
    X^{(0)} = \begin{pmatrix}
        0 &  0  & 0 \\
        0  & y_{22}^2 \alpha_{22} &  0  \\
        0  & 0  & 0
    \end{pmatrix},
\end{equation}
\begin{equation}
    X^{(1)} = \begin{pmatrix}
        0 &   y_{22}y_{12}\alpha_{22}  + y_{22}y_{11}\alpha_{12}& 0 \\
        y_{22}y_{12}\alpha_{22} + y_{22}y_{11}\alpha_{12}   & 0 &  0  \\
        0  & 0  & 0
    \end{pmatrix}, 
\end{equation}
\begin{equation}
    \resizebox{1\textwidth}{!}{$
    X^{(2)} = \begin{pmatrix}
    y_{11}^2\alpha_{11} + y_{12}^2\alpha_{22} + 2y_{11}y_{12}\alpha_{12} & 0 & 0 \\[6pt]
    0 & 2y_{21}y_{22}\alpha_{12} & y_{31}y_{22}\alpha_{12} + y_{22}y_{32}\alpha_{22} + y_{22}y_{33}\alpha_{23}  \\[6pt]
    0 & y_{31}y_{22}\alpha_{12} + y_{22}y_{32}\alpha_{22} + y_{22}y_{33}\alpha_{23}  & 0
    \end{pmatrix}$}.
\end{equation}

Using second-order perturbation theory, keeping corrections up to ${\mathcal
  O}(\epsilon^2)$ 
to the matrix and denoting the pre-factor as $m_{NH}=\frac{v^2q_Y^{-2K}\epsilon^{-4}}{2\Lambda_{\text{FN}}}$, we find that the three eigenvalues (masses) of the three lightest approximately left-handed (LH) neutrinos are 
\begin{equation} 
\begin{aligned}
    m_{\nu_3} &= m_{NH}\Bigg[
    y_{22}^2 \alpha_{22}
    + \epsilon^2 \Bigg(2\alpha_{12}y_{21}y_{22}+ \frac{(y_{12}\alpha_{22} + y_{11}\alpha_{12}\big)^2}{\alpha_{22}}
    \Bigg)
    \Bigg],\\[6pt]
    m_{\nu_2} &= \epsilon^2 m_{NH} \Bigg[
    \frac{(-\alpha_{12}^2 + \alpha_{11}\alpha_{22})y_{11}^2}{\alpha_{22}}
    \Bigg],\\[4pt]
    m_{\nu_1} &= 0. \label{NormalHierarchy}
\end{aligned}
\end{equation}
It is apparent, therefore, that for $\epsilon \ll 1$, the eigenvalues follow the NH,
$m_{\nu_1} \ll m_{\nu_2} \ll m_{\nu_3}$. Given the structure of the matrices,
$X^{(i)}$, in~\cref{Lowest order - Unchanged}, the leading order estimate for
the atmospheric mixing angle is $\theta_{23} \approx 0$ in contradiction with
the data~\cref{tab:neutrino_angles_masses}. However, the sub-leading terms in
the expansion of the angle contain ratios of independently distributed
$\mathcal{O}(1)$ coefficients and, as per the argument reproduced above from
Ref.~\cite{Allanach:2026rkb}, can have heavy 
tails, which could offset the $\epsilon$ suppression. The underlying $y_{ij}$ and
$a_{ij}$ need not necessarily be huge to achieve this. A demonstration that a
fit to the valid parameter space is natural is deferred to a numerical
scan, described below.

To obtain an IH of eigenvalues of the neutrino mass matrix, we
shall show in the next section that the condition, $a_{13} = a_{11} = 0
\implies \alpha_{22} = 0$, is sufficient. From the hierarchy in \cref{NormalHierarchy} one obtains an estimate for $\epsilon$ by matching the mass hierarchy and fixing homogeneous rational polynomials of $\mathcal{O}(1)$ numbers to also be $\mathcal{O}(1)$:
\begin{equation}\label{eq:epsilon-estimate}
    \epsilon^{-2}=\mathcal{O}(1) \times \frac{m_{\nu_3}}{m_{\nu_2}}
    \approx\mathcal{O}(1) \times \sqrt{\frac{\Delta m^2_{31}}{\Delta m^2_{21}}}
    \quad\Rightarrow\quad
    \epsilon = \Big(\mathcal{O}(1) \times \frac{7.5\times 10^{-5}}{2.5\times 10^{-3}}\Big)^{1/4} = 0.42, 
\end{equation}
where we have  replaced $\Delta m_{31}^2$ and $\Delta m_{21}^2$ by their experimental central values from \cref{tab:neutrino_angles_masses} and fixed the unknown $\mathcal{O}(1)$ factor to be unity in the final equality.
Using $\epsilon=0.42$ from \cref{eq:epsilon-estimate}, $m_{\nu_3}$ and
$m_{\nu_2}$ from \cref{NormalHierarchy}, we obtain $\Lambda_{\text{FN}} \sim \mathcal{O}(10 \ \mathrm{TeV})$ for the inferred value of $\Delta m_{31}^2$ from Table~\ref{tab:neutrino_angles_masses}. 
However, $\epsilon=0.42$ is \emph{not} a very small value; it calls into
question the accuracy and validity of the perturbation theory methods we have
been using to calculate the neutrino masses and mixing parameters. We instead
will perform a numerical diagonalisation in order to
accurately calculate the neutrino masses, mixing parameters and 
the associated likelihood from the comparison to oscillation data.

\subsection{Numerical parameter scan for NH} \label{sec:num}

To map the full parameter space and isolate regions that reproduce the observed neutrino oscillation data in~\cref{tab:neutrino_angles_masses}, we perform a global Bayesian parameter inference. Some of the uncertainties in the table are
mildly asymmetric. In our numerical work, we replace these by a symmetric
uncertainty which is the average of the two asymmetric ones, for each quantity.
Initially, gradient-based Markov Chain
Monte Carlo (MCMC) methods such as Hamiltonian Monte Carlo (HMC) were
tried, but they exhibited poor coverage across the 17-dimensional parameter
space for the NH scan (and, below, across the 15-dimensional IH scan parameter
space). Specifically, independent MCMC chains 
found posterior distributions for the parameters that were too different. 

Then we 
performed the global Bayesian parameter scan using nested sampling and the
Monte Carlo algorithm 
MLFriends~\cite{Buchner:2014qmx, Buchner_2019} as
implemented with the Python package
\texttt{UltraNest}(v4.5.0)\footnote{\url{https://johannesbuchner.github.io/UltraNest/}}
\cite{buchner2021ultranestrobustgeneral}. In the hope of encouraging robust
posterior coverage, 
the algorithm's convergence criteria required a remaining evidence threshold
of $\Delta \log Z = 0.2$, a minimum effective sample size of $20\ 000$, and a
minimum of $50\ 000$ live points, expanding dynamically to track multiple
modes. This package was used in conjunction with \texttt{NumPy}(v1.21.5) and
\texttt{SciPy}(v1.8.0). To the extent that the algorithm has converged, 
     {\tt UltraNest} gives us a sampling where the density of points is
     proportional to the posterior PDFs.

Our analytic arguments \cref{eq:epsilon-estimate} yielded the estimate
$\epsilon = 0.42$; we 
choose a uniform prior in $[0.1,0.5]$. The lower limit of $0.1$ comes
from requiring a not--too--large hierarchy between the flavon VEV
$\langle\theta\rangle$ and the mass scale suppressing the FN diagrams, $\Lambda_{\text{FN}}$.
The upper bound of $0.5$ comes from requiring the FN mechanism to have a domain
of validity.
The Majorana mass scale
$\Lambda_{\text{FN}}\sim {\mathcal O}(10)\text{~TeV}$ acts as a scale parameter, affecting
the approximately light neutrino masses through
\begin{equation}
m_{\nu_L}= -\frac{v^2}{2q_Y^{2K}\Lambda_{\text{FN}}}Y_\nu
\Sigma^{-1}A^{-1}\Sigma^{-1}Y_\nu^T.
\end{equation}
The argument that such a dimensionful scale parameter should have a
prior that is invariant under scale transformations and is non-informative
implies that we should use the Jeffreys prior
$p(\Lambda_{\text{FN}}) \propto 1/\Lambda_{\text{FN}}$
\cite{jeffreys_invariant_1946}. This is equivalent to assigning a uniform
distribution to the logarithm of this scale. Here, we choose a prior
$\log_{10}(\Lambda_{\text{FN}}/\text{GeV})\sim \mathcal{U}[1,5]$.
We note that the lower limit of this domain would imply $\langle \theta
\rangle$ below the TeV scale, at odds with the TeV-scale $Z^\prime$ premise;
however, the posterior is concentrated in the consistent
$\log_{10}(\Lambda_{\text{FN}}/\text{GeV})>3.2$ region.
To ensure perturbative control, we fix $|y_{ij}|<3$
with the priors $y_{ij} \sim \mathcal{U}[-3,3]$ and $a_{ij} \sim
\mathcal{U}[-3,3]$.

\begin{figure}[h!]
    \centering\includegraphics[width=0.7\textwidth]{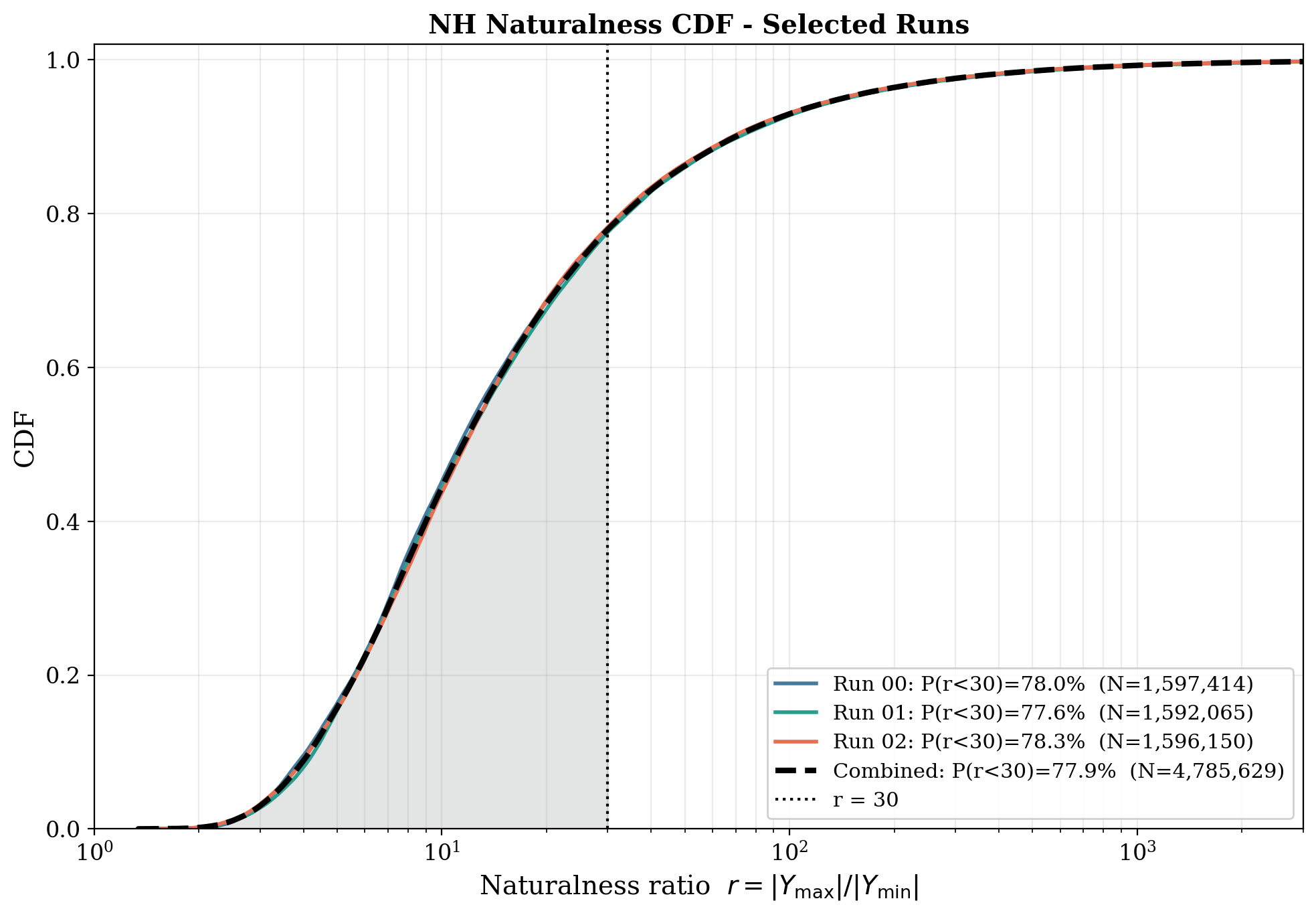}
    \caption{CDF of the spread $r$ for NH.}
    \label{fig:Normal_spread}
\end{figure}

As advertised above, we use $r$ from \cref{eq:spread} to quantify the
naturalness of the set of dimensionless Yukawa couplings in the posterior PDF.
Across three independent nested sampling runs, each with $N_{\mathrm{live}} =
50\ 000$, the model exhibits a naturalness ratio CDF at the value of 30 as 
\begin{equation} \label{NO:CDFPercent}
    P(r < 30) = 78.0 \pm 0.4\%,
\end{equation} 
with the standard deviation estimated from the run-to-run variance. The CDF of
$r$ across all three runs is shown in \cref{fig:Normal_spread}, with the three
profiles showing close agreement. 
We note here that under a $\mathcal{U}[-3,3]$ PDF for 9 couplings,
$P(r<30)=0.76$.
We deduce that the fit is therefore of comparable naturalness to
this PDF. 

\begin{figure}[h!]
    \centering
    \includegraphics[width=0.9\textwidth]{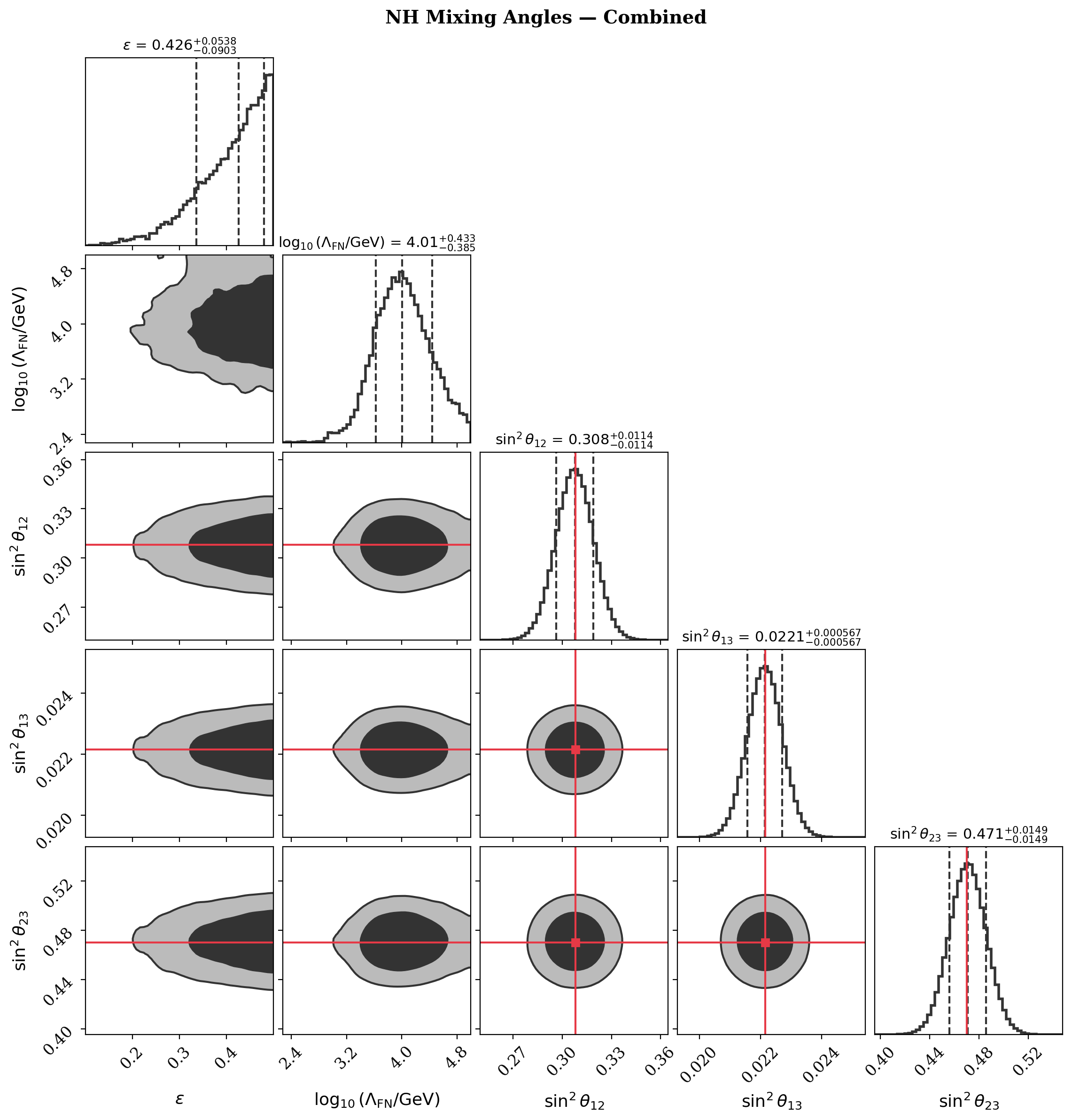}
    \caption{Posterior probability distributions of the mass scale, FN parameter and the mixing angles (NH).}
    \label{fig:CornerPlot_NO_angles}
\end{figure}

\begin{figure}[h!]
    \centering
    \includegraphics[width=0.9\textwidth]{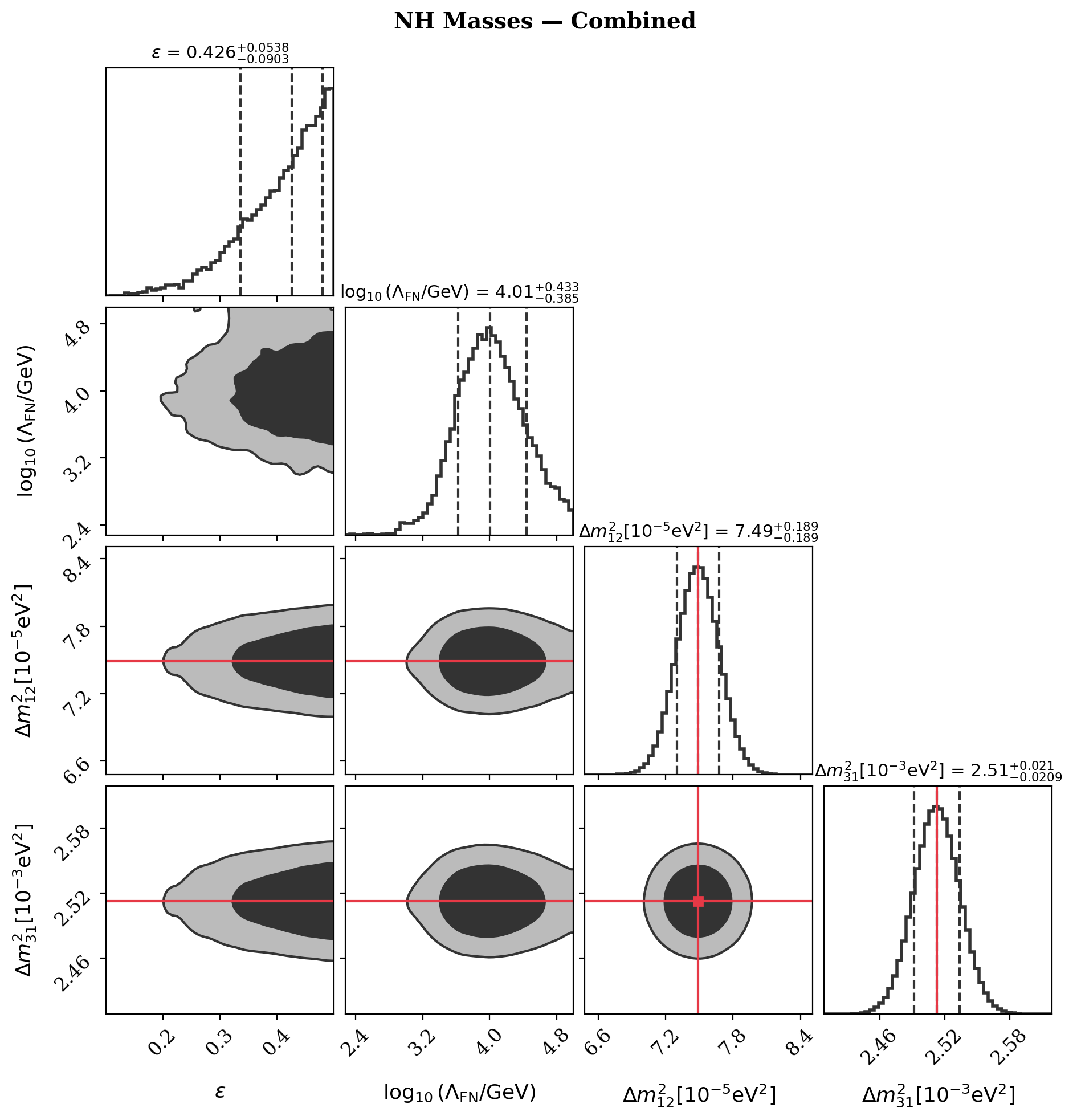}
    \caption{Posterior probability distributions of the mass scale, FN parameter and the mass squared differences (NH). The solid contours correspond to the 68$\%$ and 95$\%$ Bayesian credibility intervals.}
    \label{fig:CornerPlot_NO_masses}
\end{figure}

The posterior distributions of the predicted angles, heavy neutrino mass scale
and the mass-squared differences are presented in
\cref{fig:CornerPlot_NO_angles} and \cref{fig:CornerPlot_NO_masses}.
The posterior shown in the $(\epsilon, \log_{10}
(\Lambda_{\text{FN}}/\text{GeV)})$ plane is predominantly 
connected, and concentrated near $\epsilon \sim 0.42$ and $\Lambda_{\text{FN}} \sim \mathcal{O}(10) \ \text{TeV}$, as expected from our
qualitative estimates in \cref{eq:epsilon-estimate}. The posterior means of the
three mixing angles and the two mass squared differences are given in
\cref{tab:NO_Posteriors}. The global
evidence is $\log Z_{NH} = -26.52 \pm 0.02$.  

\begin{table}[h!]
    \centering
    \begin{tabular}{|c|c|c|}
      \hline
        Observable & Posterior (mean $\pm 1 \sigma$) & Experimental (mean $\pm$ 1$\sigma$) \\
        \hline
        \hline
        $\Delta m_{21}^2$/eV$^2$ & $(7.49 \pm 0.19)\times 10^{-5}$ & $(7.49 \pm 0.19)\times 10^{-5}$  \\
        $\Delta m_{31}^2$/eV$^2$ & $(2.51 \pm 0.02)\times 10^{-3}$ & $(2.51 \pm 0.02)\times 10^{-3}$ \\
        $\sin^2\theta_{12}$ & $0.3077 \pm 0.0115$ & $0.3080 \pm 0.0115$  \\
        $\sin^2\theta_{13}$ & $0.0221 \pm 0.0006$ & $0.0221 \pm 0.0006$ \\
        $\sin^2\theta_{23}$ & $0.4708 \pm 0.0150$ & $0.4700 \pm 0.0150$  \\
        \hline
    \end{tabular}
    \caption{Posterior means and 1$\sigma$ standard deviations for the Normal
      Hierarchy fit 
      under Nested Sampling. The experimental values and uncertainties are
      taken from {\tt NuFIT 6.0} (2024) global fits with Super Kamiokande (SK) data~\cite{Esteban:2024eli}.}
    \label{tab:NO_Posteriors}
\end{table}

It is apparent from \cref{tab:NO_Posteriors} that the sampled posterior means of
neutrino mass squared differences and mixing angles are all very similar to
the experimental inputs. This is no surprise since our system is
under-constrained; it has more parameters than there are measured experimental
observables.

\subsection{Analytic approximation for IH} \label{sec:IH}
As established in~\cref{sec:anal}, the leading-order mass matrix $X^{(0)}$ is
diagonal with a single non-zero entry, $y_{22}^2 \alpha_{22}$, at
$\mathcal{O}(\epsilon^0)$. In the language of non-degenerate
Rayleigh-Schr\"odinger perturbation theory, $\alpha_{22}$ acts as the
unperturbed energy gap ($\Delta$) separating the heaviest state from the
massless subspace. The mixings induced by $X^{(1)}$ and $X^{(2)}$ act as small
off-diagonal perturbations ($V$). When the gap is parametrically larger than
the perturbations ($|\Delta| \gg |V|$), non-degenerate perturbation theory
applies. This yields the widely split, hierarchical spectrum characteristic of
the NH\@. However, IH requires the two heaviest eigenvalues to be degenerate
at leading order $m_{\nu_3} \ll m_{\nu_1} \approx m_{\nu_2} $. To transition
to the IH, one must force the unperturbed energy gap to close ($|\Delta|
\lesssim |V|$), thereby intentionally breaking the non-degenerate expansion
and driving a level-crossing. This is achieved by imposing $\alpha_{22} \ll
1$, which shifts the system strictly into the domain of degenerate
perturbation theory. 

A parametric suppression of the entry $\alpha_{22} \sim
\mathcal{O}(\epsilon^2)$, so that $a_{11} \sim \mathcal{O}(\epsilon^2)$ and
$a_{13} \sim \mathcal{O}(\epsilon)$, is also sufficient to generate the
suppression required to move between the regimes and can be dynamically realised
by the additional clockwork mechanism described in~\cref{Sec:Clockwork}. We
shall work with 
exact texture zeros for clarity in our analytic arguments.
(Ref.~\cite{Frampton:2002yf} found seven neutrino mass textures with
two independent zeroes that were acceptable with the 2002 dataset.)  
\begin{equation} \label{eqn:IOtexture}
    M = \Lambda_{\text{FN}}
    \begin{pmatrix}
    0&\epsilon^{3}a_{12}&0\\
    
    \epsilon^{3} a_{12} & \epsilon^{4} a_{22}& \epsilon^{2} a_{23}\\
    
    0 & \epsilon^{2}a_{23}&a_{33} 
    \end{pmatrix}.
\end{equation}

We expand the matrix $m_\nu$ as a series in $\epsilon$, similar to~\cref{eqn:mnu_expansion}, after factoring out some common factors, 
\begin{equation}
    m_\nu = \frac{v^2 q_Y^{-2K}}{2\Lambda_{\text{FN}}} \epsilon^{-3} \left[\tilde X^{(0)} + \epsilon \tilde
    X^{(1)} + \epsilon^2 \tilde X^{(2)} + \mathcal{O}(\epsilon^3)\right] 
\end{equation}
where the matrices $\tilde X^{(i)}$ have the form
\begin{equation}
    \tilde X^{(0)} = \begin{pmatrix}
        0 &  \times & 0 \\
        \times  & 0  &  0  \\
        0  & 0  & 0
    \end{pmatrix}, \,\,\,\,  \tilde X^{(1)} = \begin{pmatrix}
        \times & 0 & 0 \\
        0 &  \times &\times  \\
        0  & \times & 0
    \end{pmatrix}, \,\,\,\, 
    \tilde X^{(2)} = \begin{pmatrix}
    0 & \times & \times \\
    \times & 0& 0  \\
    \times& 0&0 
    \end{pmatrix}
\end{equation}
We once again collect the overall coefficient setting the mass scale $m_{IH} := \frac{v^2 q_Y^{-2K}}{2\Lambda_{\text{FN}}}\epsilon^{-3}$ as in \cref{sec:anal}. We obtain
mass eigenvalues\footnote{We fix the neutrino fields' phases here such that all masses $m_{i} \geq 0$.} (masses) 

\begin{align} 
    m_{\nu_1} &= m_{IH} \left(A + \epsilon B+ \epsilon^2
    C+{\mathcal{O}}(\epsilon^3) \right) \nonumber \\
    m_{\nu_2} &= m_{IH} \left(A - \epsilon B+  \epsilon^2
    C+{\mathcal{O}}(\epsilon^3) \right)\label{eqn:IHmasses} \\
    m_{\nu_3} &= m_{IH} {\mathcal{O}}(\epsilon^3), \nonumber 
\end{align}
where $A$ and $B$ are
homogeneous rational functions of the $a_{ij}$ and $y_{ij}$.
\cref{eqn:IHmasses} corresponds to an IH i.e.,\ $m_{\nu_3} \ll m_{\nu_1}
\approx m_{\nu_2}$, as
expected. 
\cref{eqn:IHmasses} implies
$\Delta m_{21}^2 / |\Delta m_{32}^2| \approx 4 (B/A) \epsilon$; for $B/A=1$,
matching the measured value of $0.030$
(from central values in \cref{tab:neutrino_angles_masses}) would require
$\epsilon \approx 0.0075$, which is too small
to reproduce the experimentally inferred
mixing angles. Consequently, the ratio $A/B$ would need to be large to fit the
IH spectrum. This is not as
unlikely as one might imagine due to heavy tails in
ratios of order unity dimensionless couplings~\cite{Allanach:2026rkb}. We
shall implement the numerical nested sampler to fit 
the parameters, $y_{ij},a_{ij},\Lambda_{\text{FN}}$ and $\epsilon$, with a
likelihood function calibrated against the IH angles and mass squared
differences from \cref{tab:neutrino_angles_masses}.

\subsection{Numerical scan for IH}

We once again perform a global Bayesian parameter inference to map the parameter space that reproduces the observed mixing angles as in \cref{tab:neutrino_angles_masses}, using the same Python
modules and configurations (\texttt{UltraNest}, $N_{\mathrm{live}} = 50\ 000$, $\Delta \log Z = 0.2$ and an effective sample size of $20\ 000$). 
We use the same priors as in the NH case, except that we fix $a_{11}=a_{13}=0$.

\cref{eq:spread} again gives us our measure of naturalness $r$ for the 9
dimensionless neutrino Yukawa couplings.
Over the posterior, we obtain
\begin{equation} \label{IO:CDFPercent}
    P(r<30) = 69 \pm 5 \%
\end{equation}
with the standard deviation again estimated from the run-to-run
variance. Compared to the case of the NH, the
sampling algorithm
finds it harder to converge in the IH case, which results in a higher variance.
Similar to the NH case, the fit is 
of comparable naturalness to (but slightly less natural than) the
$\mathcal{U}[-3,3]$ PDF for 9 couplings, where $P(r<30)=0.76$.

The physics observables, however, such as the
three mixing angles and the mass-squared differences, agree with the
{\tt NuFIT 6.0} \cite{Esteban:2024eli} best fit values within $1 \sigma$. This
is presented in \cref{tab:IO_Posteriors} and in the corner plots \cref{fig:CornerPlot_IO_angles} and \cref{fig:CornerPlot_IO_masses}. Given the CDF percentage
uncertainties, we can say that in the \model, both mass orderings (NH or IH) are reasonably natural, with around two thirds
of the posterior volume exhibiting our naturalness criterion of $r<30$. The
global evidence in the IH case is $\log Z_{IH} = -29.07 \pm 0.03$. The
Bayes factor is then $\exp (\log Z_{NH} - \log Z_{IH}) = 12.81$ in favour of
NH, which is strong evidence in support of NH over
IH~\cite{jeffreys_theory_1998, lee_bayesian_2014} within the \model{}.
Each evidence uses the corresponding NuFIT column
of~\cref{tab:neutrino_angles_masses}, so the 
ratio omits the data's own ordering preference:
there already exists a
preference for NH in the NuFIT global analyses, since it has a $\chi^2$ value
which is 6.1 units lower than that of the IH. 

\begin{figure}[h!]
    \centering\includegraphics[width=0.7\textwidth]{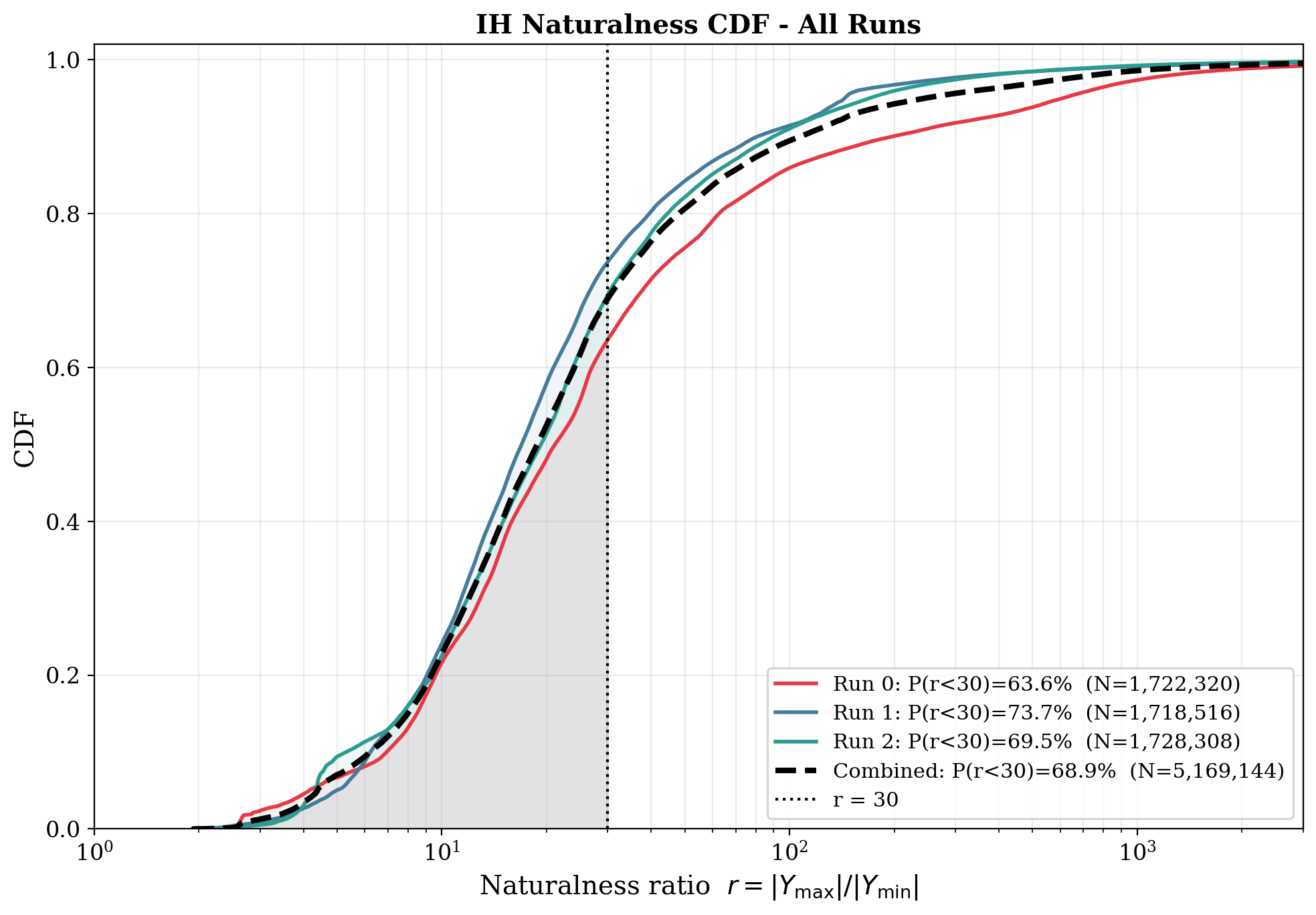}
    \caption{CDF of the spread $r$ for IH.}
    \label{fig:Inverted_spread}
\end{figure}
\begin{figure}[h!]
    \centering
    \includegraphics[width=0.9\textwidth]{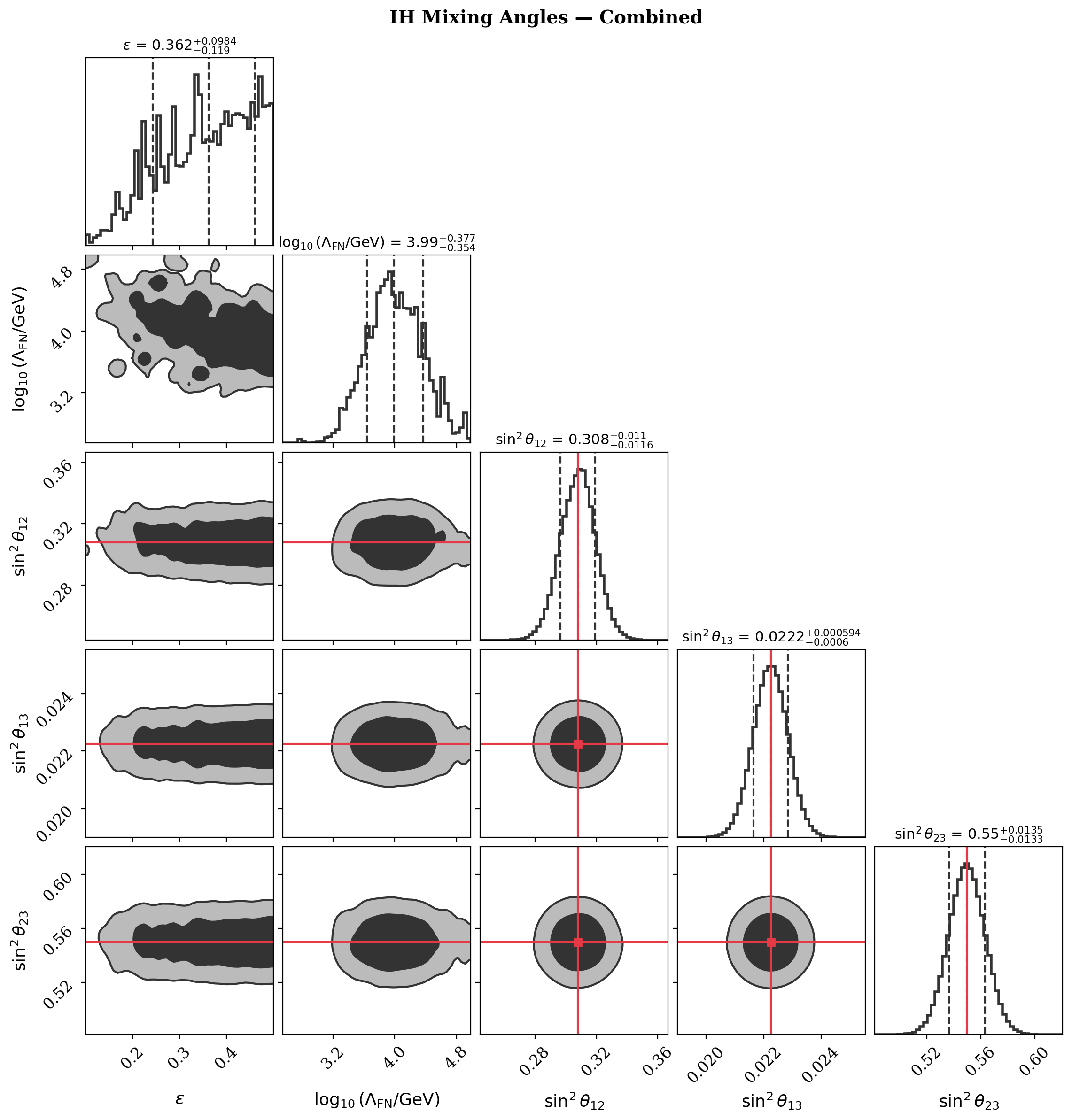}
    \caption{Posterior probability distributions of the mass scale, FN parameter and the mixing angles (IH).}
    \label{fig:CornerPlot_IO_angles}
\end{figure}
\begin{figure}[h!]
\centering
    \includegraphics[width=0.9\textwidth]{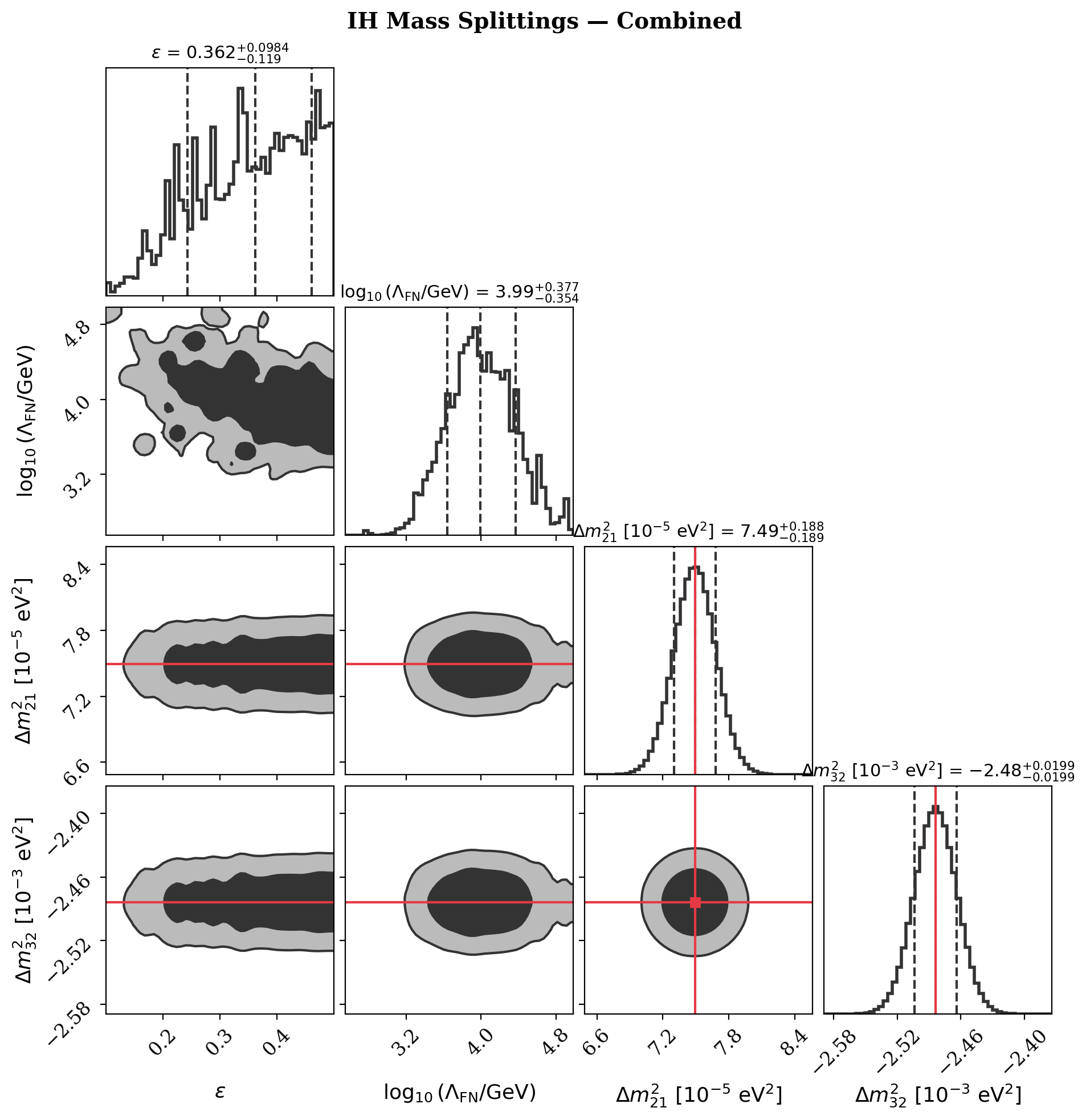}
    \caption{Posterior probability distributions of the mass scale, FN
      parameter and the mass squared differences (IH).}
    \label{fig:CornerPlot_IO_masses}
\end{figure}

\begin{table}[h!]
    \centering
    \begin{tabular}{|c|c|c|}
      \hline
        Observable & Samples (mean $\pm$ 1$\sigma$) & Experimental (mean $\pm$ 1$\sigma$) \\
        \hline
        \hline
        $\Delta m_{21}^2$/eV$^2$ & $(7.49 \pm 0.19)\times 10^{-5}$ & $(7.49 \pm 0.19)\times 10^{-5}$  \\
        $\Delta m_{32}^2$/eV$^2$ & $(-2.48 \pm 0.02)\times 10^{-3}$ & $(-2.48 \pm 0.02)\times 10^{-3}$  \\
        $\sin^2\theta_{12}$ & $0.3081 \pm 0.0114$ & $0.3080 \pm 0.0115$ \\
        $\sin^2\theta_{13}$ & $0.0222 \pm 0.0006$ & $0.0223 \pm 0.0006$ \\
        $\sin^2\theta_{23}$ & $0.5497 \pm 0.0134$ & $0.5500 \pm 0.0135$ \\
      \hline
    \end{tabular}
    \caption{Posterior means and 1$\sigma$ standard deviations for the IH 
      fit under Nested Sampling. The experimental values and uncertainties are
      taken from {\tt NuFIT 6.0} (2024) global fits (with SK data) \cite{Esteban:2024eli}.}
    \label{tab:IO_Posteriors}
\end{table}

\section{Conclusions}\label{sec:conc}
We have shown that
the \model\ can explain the small neutrino masses and
sizeable leptonic mixing angles that agree with those inferred from
measurements of neutrino oscillation phenomena. 
The $U(1)_X$ gauge symmetry in the \model\ is successfully and
simultaneously put to three uses: to better fit measurements involving the
\bsll\ transition, to explain some of the hierarchies between neutrino masses
and to suppress CLFV\@. 
Successful neutrino mass patterns require an
additional suppression mechanism such as clockwork for both the NH and IH
cases of neutrino spectrum\footnote{We note that in
Refs.~\cite{Alonso:2018bcg}, in a different setting, the clockwork mechanism was
applied directly to the quark flavour puzzle and it was shown that TeV-scale `gears'
could be made to work for quark masses and mixing angles without the need for FN at all. 
However, we are instead modelling the 
neutrino masses and mixing angles and we are using a hybrid FN-clockwork
approach. All charged fermion masses were clockworked in
Ref.~\cite{AbreudeSouza:2019ixc}, with random order unity fundamental
dimensionless couplings.}.
Some mildly large but acceptable
factors are required between the dimensionless couplings in order to acquire
leptonic mixing angles that agree with those inferred from neutrino
oscillation data\footnote{In Ref.~\cite{Vempati:2025dow}, it was shown how
randomness in disordered fermion theory spaces can lead to
anarchical neutrino mixing either through Anderson localisation or through a
GIM-like cancellation.}. These
mild factors are quantified in the spread $r$, whose distributions are
displayed in Figs.~\ref{fig:Normal_spread} and~\ref{fig:Inverted_spread}, respectively.
For the IH case, two entries in the right-handed Majorana neutrino mass matrix
require further suppression, possibly by an 
additional family-dependent clockwork mechanism, as exemplified in
Appendix~\ref{Sec:Clockwork}.  

Assuming that the lightest neutrino has a negligible mass,
the sum of the three
approximately left-handed light neutrino masses is around 0.10~eV in the IH
case, in tension 
with the more aggressive DESI+CMB 
combinations of cosmological
fit~\cite{collaboration_desi_2025}.
The light neutrino mass matrix then predicts the mass parameter relevant for
neutrinoless double beta decay
$m_{\beta\beta}=18-48$~meV -- within reach of the experiments
LEGEND-1000/nEXO~\cite{collaboration_search_2023,collaboration_legend-1000_2021,collaboration_nexo_2022}. Thus,
the Plan B IH Type I seesaw hypothesis would be
tested in the not-too-distant future. The Plan B NH counterparts of the cosmology
predictions ($\sum m_\nu \approx 0.06$ eV and $m_{\beta\beta}$ of a few meV)
predict that 
neutrinoless double beta decay would not be seen in the next round of
experiments. 

The approximately RH neutrinos have masses ranging from $\epsilon^3
\Lambda_{\text{FN}}$ up to $\Lambda_{\text{FN}}\sim {\mathcal O}(10)$~TeV, meaning
that they could participate in $Z^\prime$ phenomenology. If the $Z^\prime$ were
massive enough, $Z^\prime \rightarrow \nu_{R_i} \nu_{R_j}$ decays would
modify the $Z^\prime$ width and branching ratios. This could in principle
change the quoted lower bounds upon
$M_{Z^\prime}$~\cite{Allanach:2024jls,Allanach:2026yst} from searches at colliders.
The total $Z^\prime$ width would increase by a factor 1.18, were the
$\nu_{R_{i,j}}$ to be massless, and so this places an upper bound on how much the
branching ratio into di-muon pairs could \emph{decrease}: by a factor
$1/1.18=0.84$. Current 
LHC bounds on the \model\ $Z^\prime$ are dominated by
searches for resonant di-muon pairs and are sensitive to the cross-section
times branching ratio. Fig.~3a of Ref.~\cite{Allanach:2024jls} shows that such
a change will not have a large effect upon the lower mass bound.

One limitation of our analysis is the
restriction to real parameters only: it would be interesting to complexify and
to make leptogenesis potentially workable. In the IH case, the
heavy neutrino mass spectrum (with masses of order $\Lambda_{\text{FN}},\ \epsilon^{3}\Lambda_{\text{FN}},\
\epsilon^3 \Lambda_{\text{FN}}$) is of interest here.  

It is clear that the \model\ would need additional model building in order to
extend the FN mechanism to fully explain all of the observed patterns in the
charged fermionic fields' masses and mixing angles in detail. One requires an
extra symmetry that will enforce an approximate $U(2)_q \times U(2)_u \times
U(2)_d$ global symmetry and make the upper left-hand block of $Y_{U,D}$
(\ref{chferms}) zero, at the level of the renormalisable terms in the
Lagrangian density. 
We envisage this to follow the flavour deconstruction
programme~\cite{Bordone:2017bld,Greljo:2024ovt}, where the details of the
lighter charged fermion masses and mixing angles will be explained further
into the ultra-violet r\'egime than the multi-TeV scales modelled here
by augmenting with further family gauge symmetries.

Flavour deconstruction has recently attracted attention as regards neutrino
masses in other settings: since neutrino masses and mixing angles look
somewhat anarchic compared to the charged fermions of the
SM~\cite{Hall:1999sn,deGouvea:2012ac}, it was not \emph{a priori} obvious that
it would explain the patterns inferred from neutrino oscillation
measurements. However, some generalities were considered in
Ref.~\cite{Greljo:2024ovt}, where various models with flavour deconstructed
$U(1)_R \times U(1)_{B-L}$ type symmetries at the heart of them were shown to
yield enough neutrino anarchy. More recently, it was shown how extending this
to flavour deconstructed $SU(2)_L \times U(1)_R \times U(1)_{B-L}$ gauge
symmetry yields flavour anarchy with a particularly low (but still TeV-scale)
spontaneous symmetry breaking scale~\cite{Isidori:2025rci}.
In Ref.~\cite{FernandezNavarro:2025zmb}, it was shown how flavour
deconstruction in terms of a 
tri-hypercharge model can result in a successful neutrino mass
pattern and lepton mixing that agrees with that inferred from neutrino
oscillation experiments.
These three works did not attempt to simultaneously improve agreement of the
predictions of $B$ observables with measurements, unlike the \model. 
It has already been pointed out in Ref.~\cite{Bonilla:2017lsq} that
$U(1)_{B_3-L_\mu}$ (in our normalisation) gauge symmetry, which was proposed
to explain earlier 
neutral current $B$-anomaly measurements, would yield a Type I seesaw mechanism, although no detailed study of the neutrino sector was performed. 

The \model\ has the following to commend it within its preferred parameter space: neutrino masses and mixing angles
that quantitatively fit data well with order unity
fundamental Yukawa 
couplings once a clockwork mechanism is additionally employed, a few features of the charged fermion mass spectrum are explained
qualitatively and fits to measurements involving $B$ mesons significantly improve
upon those of the SM.

\section*{Acknowledgements}
This work was partially supported by STFC HEP Consolidated grant
ST/X000664/1. We thank the Cambridge Pheno Working Group for discussions. 

\appendix
\section{An additional clockwork mechanism for IH Majorana mass matrix entry suppression} \label{Sec:Clockwork}
A possible way to achieve the Majorana mass elements $a_{11},a_{13} \ll 1$
suppression required in the IH case of 
\cref{sec:IH} is via an additional clockwork mechanism
\cite{Kaplan:2015fuy,Giudice:2016yja,Choi:2015fiu}. This is different from the
setup employed in~\cref{sec:anal} in that flavour-dependent groups are introduced here.
The focus, primarily, will be on the bare Majorana matrix $A$ given in \cref{eqn:MajoranaMatrixFact}. This can be extended easily to the case with mediators to create the FN suppression. 

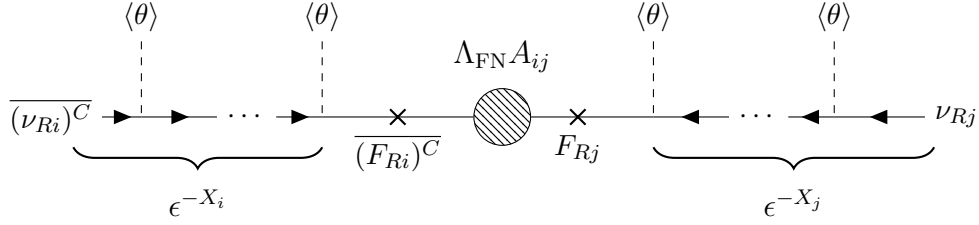
\begin{figure}[t] 
    \centering
    \begin{tikzpicture}
        \begin{feynman}
            \vertex (nuL) {$\overline{(\nu_{Ri})^C}$};
            
            \vertex [right=1.2cm of nuL] (v1);
            \vertex [above=1.0cm of v1] (th1) {$\langle \theta \rangle$};
            
            \vertex [right=1.0cm of v1] (dotsL) {$\cdots$};
            
            \vertex [right=1.0cm of dotsL] (v2);
            \vertex [above=1.0cm of v2] (th2) {$\langle \theta \rangle$};

            \vertex[right=1.0cm of v2] (F1);
            \vertex[below=0.1cm of F1] (F1label) {$\overline{(F_{Ri})^C}$};
            
            \vertex [blob, right=1.0cm of F1] (CW) {};

            \vertex[right=1.0cm of CW] (F2);
            \vertex[below=0.1cm of F2] (F2label) {$F_{Rj}$};
            \vertex [right=1.0cm of F2] (v3);
            \vertex [above=1.0cm of v3] (th3) {$\langle \theta \rangle$};
            
            \vertex [right=1.0cm of v3] (dotsR) {$\cdots$};
            
            \vertex [right=1.0cm of dotsR] (v4);
            \vertex [above=1.0cm of v4] (th4) {$\langle \theta \rangle$};
            
            \vertex [right=1.2cm of v4] (nuR) {$\nu_{Rj}$};
            
            \diagram*{
                (nuL) -- [fermion] (v1) -- [fermion] (dotsL) -- [fermion] (v2) --  (CW),
                (th1) -- [scalar] (v1),
                (th2) -- [scalar] (v2),
                
                (nuR) -- [fermion] (v4) -- [fermion] (dotsR) -- [fermion] (v3) -- (CW),
                (th3) -- [scalar] (v3),
                (th4) -- [scalar] (v4),
            };

            \node [cross out, draw=black, thick, minimum size=5pt, inner sep=0pt] at (F1) {};
            \node [cross out, draw=black, thick, minimum size=5pt, inner sep=0pt] at (F2) {};
        \end{feynman}
        
        \draw [decorate, decoration={brace, amplitude=8pt, mirror}, thick] 
            ([xshift=0.3cm, yshift=-0.4cm]nuL.center) -- ([yshift=-0.4cm]v2.center) 
            node [midway, below=12pt] {\large $\epsilon^{-X_i}$};
            
        \draw [decorate, decoration={brace, amplitude=8pt, mirror}, thick] 
            ([yshift=-0.4cm]v3.center) -- ([xshift=-0.3cm, yshift=-0.4cm]nuR.center) 
            node [midway, below=12pt] {\large $\epsilon^{-X_j}$};
            
        \node at ([yshift=0.8cm]CW) {\large $\Lambda_{\text{FN}} A_{ij}$};
                
    \end{tikzpicture}
    \caption{Feynman diagram illustrating integrated FN and clockwork mechanisms generating the effective Majorana operator $\Lambda_{\text{FN}} \epsilon^{-X_i-X_j} \overline{(\nu_{Ri})^C} \nu_{Rj}$ ($-X_i>0$ here for the \model\ charges). The external RH neutrinos acquire FN suppression factors $\epsilon^{-X_{i}}$ and $\epsilon^{-X_j}$ via successive insertions of the flavon field $\langle\theta\rangle$. The central blob represents the heavy Majorana mass operator $\Lambda_{FN}\overline{(F_{Ri})^C}F_{Rj}$, generated by integrating out the $U(1)_X$-neutral clockwork mediators represented by the blob.}
    \label{fig:IntCWinFN}
\end{figure}

First, looking at the factorisation property of $M$,
\begin{equation}
    \label{eqn:MajoranaMatrixFact2}
M = \Lambda_{\text{FN}}\Sigma A \Sigma, \qquad 
    \Sigma = \begin{pmatrix}
    \epsilon^{1} & 0 & 0 \\
     0 & \epsilon^2 & 0 \\
     0 & 0 & 1 \\
    \end{pmatrix}, \qquad
A =     \begin{pmatrix}
        a_{11} & a_{12}  & a_{13} \\
        a_{12} & a_{22}  & a_{23} \\
        a_{13} & a_{23} & a_{33} 
    \end{pmatrix},
\end{equation} 
we can depict this as shown in \cref{fig:IntCWinFN}. The inter-mediator particles $F_{i}$ have zero $U(1)_X$ charge, but possess the same global symmetries as the corresponding $\nu_i$.
We assume that the fields $\nu_1, \nu_2,\nu_3$ (correspondingly
$F_1,F_2,F_3$), have a negative unit charge $(-1,0,0), (0,-1,0),$ and $(0,0,-1)$ under global symmetry group $U(1)_A \times U(1)_B \times U(1)_C$. This forbids the Majorana mass matrix
entirely. To reintroduce the required couplings, we add spurions as before to
the UV spectrum, with the charges $(1,\ 1,\ 0), (0,\ 1,\ 1),$ $(0,\ 2,\ 0),$ and $(0,\ 0,\ 2)$. 
With this, we can write down the Lagrangian for direct couplings ($a_{22}$) with heavy gears on a mass scale of $\mu' \sim \Lambda_{\text{FN}}$ and $q>1$,
\begin{table}[h]
    \centering
    \renewcommand{\arraystretch}{1.3}
    \begin{tabular}{@{} l c c c @{}}
    \toprule
    \textbf{Field} & \textbf{Lorentz Type} & \textbf{Charge: $U(1)_A \times U(1)_B \times U(1)_C$} & $U(1)_X$ \\
    \midrule
    $\nu_{R1}$ & RH Fermion & $(-1, 0, 0)$ & $-1$ \\
    $\nu_{R2}$ & RH Fermion & $(0, -1, 0)$ & $-2$\\
    $\nu_{R3}$ & RH Fermion & $(0, 0, -1)$ & $0$\\
    \midrule
    $F_{R1}$ & RH fermions & $(-1, 0, 0)$ & $0$\\
    $F_{R2}$ & RH Fermions & $(0, -1, 0)$& $0$ \\
    $F_{R3}$ & RH Fermions & $(0, 0, -1)$ & $0$\\
    \midrule
    $\psi_i,\ \chi_i$ & Vector-like Fermions & $(0, 1, 0)$ & $0$\\
    \midrule
    $\Phi_{22}$ (for $a_{22}$) & Scalar Spurion & $(0, 2, 0)$ & $0$\\
    $\Phi_{33}$ (for $a_{33}$) & Scalar Spurion & $(0, 0, 2)$ & $0$\\
    $\Phi_{12}$ (for $a_{12}$) & Scalar Spurion & $(1, 1, 0)$ & $0$\\
    $\Phi_{23}$ (for $a_{23}$) & Scalar Spurion & $(0, 1, 1)$ & $0$\\
    \bottomrule
    \end{tabular}
    \caption{Field content and flavour charge assignments for the clockwork texture zero mechanism. The absence of spurions with charges $(2,0,0)$ and $(1,0,1)$ guarantees the required zeros at $a_{11}$ and $a_{13}$.}
    \label{tab:clockwork_charges}
\end{table}
\begin{equation} \label{eqn:IOCMdirect}
    \mathcal{L} = a_{0} \overline{(F_{2R})^C} \langle\Phi_{22}\rangle \psi_{R0} + a_1 \mu' \overline{\psi_{L1}} F_2 - \mu' q(\overline{\psi_{L0}}\psi_{R0} + \overline{\psi_{L1}}\psi_{R1}) + \mu'  \overline{\psi_{L0}} \psi_{R1} + H.c.,
\end{equation}
where we have used the spurion field with charge $(0,2,0)$. One can now do the
same for the term $a_{33}$, using the spurion with charge $(0,\ 0,\ 2)$. 

For the cross terms, we write,
\begin{equation} \label{eqn:IOCMdiagonal}
    \mathcal{L} = a_{0}' \overline{(F_{1R})^C} \langle \Phi_{12}\rangle \chi_{R0} + a_1' \mu '\overline{\chi_{L1}} F_2  - \mu' q(\overline{\chi_{L0}}\chi_{R0} + \overline{\chi_{L1}}\chi_{R1}) + \mu'  \overline{\chi_{L0}} \, \chi_{R1} + H.c.,
\end{equation}
where we have used the spurion with charge $(1,1,0)$. We can do the same
with the coupling of $\nu_2$ to $\nu_3$, where we now use the spurion with
charge $(0,\ 1, 1)$. Now, importantly, since there exists no spurion with
charge $(1,\ 0,\ 1)$ or $(2,\ 0,\ 0)$, there can be no direct coupling as written in
the Lagrangians \cref{eqn:IOCMdirect} or
\cref{eqn:IOCMdiagonal}. Consequently, the effective couplings $a_{11}$ and
$a_{13}$ are generated only at higher orders in the spurion expansion
(e.g.\ via mixing through the $\nu_2$ sector), resulting in the suppression
factors $q^{-m}$ and $q^{-n}$, respectively. Using the equations of motion, we
can write down an effective Majorana mass matrix of the form
\begin{equation}
    \Lambda_{\text{FN}}A = (\Lambda_{\text{FN}} q^{-1})
    \begin{pmatrix}
    q^{-m} a_{11}&a_{12}&q^{-n} a_{13}\\
    
    a_{12} &  a_{22}&  a_{23}\\
    
    q^{-n} a_{13} & a_{23}&a_{33} 
    \end{pmatrix}.
\end{equation}
Here, the overall factor of $q^{-1}$ that arises from the spurion couplings as
given in \cref{eqn:IOCMdirect} and \cref{eqn:IOCMdiagonal} has been factored
out of the matrix to multiply the mass scale $\Lambda_{\text{FN}}$. The couplings $a_{11}$ and
$a_{13}$ come with extra factors of $q^{-1}$, where $n,m \geq 1$, arising from
indirect (through the $\nu_2$ sector) couplings between $\nu_1$ to $\nu_3$ and
$\nu_1$ to $\nu_1$. For sufficiently large values of $q$, the couplings
proportional to $a_{11}, a_{13}$
render themselves 
negligible
relative to the dominant terms, thereby 
yielding the
requisite matrix structure characterised by the texture zeros delineated in \cref{eqn:IOtexture}. 

\bibliographystyle{JHEP-2}
\bibliography{article}

\end{document}